\documentclass[twocolumn]{aastex63}

\usepackage{amsmath}
\usepackage{color}
\usepackage{graphicx}
\usepackage[caption=false]{subfig}
\usepackage{tikz}
\usetikzlibrary{shapes.geometric, arrows}
\usepackage{multirow}
\usepackage{enumitem}

\shorttitle{Ensuring robustness in global 21-cm analysis}
\shortauthors{Bassett et al.}

\begin{document}

\title{Ensuring Robustness in Training Set Based Global 21-cm Cosmology Analysis}

\correspondingauthor{Neil Bassett}
\author[0000-0001-7051-6385]{Neil Bassett}
\affiliation{Center for Astrophysics and Space Astronomy, Department of Astrophysical and Planetary Sciences, University of Colorado Boulder, CO 80309, USA}

\author{David Rapetti}
\affiliation{NASA Ames Research Center, Moffett Field, CA 94035, USA}
\affiliation{Research Institute for Advanced Computer Science, Universities Space Research Association, Mountain View, CA 94043, USA}
\affiliation{Center for Astrophysics and Space Astronomy, Department of Astrophysical and Planetary Sciences, University of Colorado Boulder, CO 80309, USA}

\author{Keith Tauscher}
\affiliation{Center for Astrophysics and Space Astronomy, Department of Astrophysical and Planetary Sciences, University of Colorado Boulder, CO 80309, USA}
\affiliation{Department of Physics, University of Colorado, Boulder, CO 80309, USA}

\author{Jack~O.~Burns}
\affiliation{Center for Astrophysics and Space Astronomy, Department of Astrophysical and Planetary Sciences, University of Colorado Boulder, CO 80309, USA}

\author{Joshua J. Hibbard}
\affiliation{Center for Astrophysics and Space Astronomy, Department of Astrophysical and Planetary Sciences, University of Colorado Boulder, CO 80309, USA}

\email{Neil.Bassett@colorado.edu}

\begin{abstract}

We present a methodology for ensuring the robustness of our analysis pipeline in separating the global 21-cm hydrogen cosmology signal from large systematics based on singular value decomposition (SVD) of training sets. We show how traditional goodness-of-fit metrics such as the $\chi^2$ statistic that assess the fit to the full data may not be able to detect a suboptimal extraction of the 21-cm signal when it is fit alongside one or more additional components due to significant covariance between them. However, we find that comparing the number of SVD eigenmodes for each component chosen by the pipeline for a given fit to the distribution of eigenmodes chosen for synthetic data realizations created from training set curves can detect when one or more of the training sets is insufficient to optimally extract the signal. Furthermore, this test can distinguish which training set (e.g. foreground, 21-cm signal) needs to be modified in order to better describe the data and improve the quality of the 21-cm signal extraction. We also extend this goodness-of-fit testing to cases where a prior distribution derived from the training sets is applied and find that, in this case, the $\chi^2$ statistic as well as the recently introduced $\psi^2$ statistic are able to detect inadequacies in the training sets due to the increased restrictions imposed by the prior. Crucially, the tests described in this paper can be performed when analyzing any type of observations with our pipeline.

\end{abstract}


\section{Introduction}

Arising from the hyperfine splitting caused by the interaction of the magnetic moments of the proton and electron in neutral hydrogen, the 21-cm transition provides a unique probe of the universe's history between the Cosmic Microwave Background (CMB) and reionization. Radiation of wavelength 21-cm, corresponding to a rest frequency of 1420 MHz, produced during this time period is redshifted to frequencies of $\sim 10 - 200$ MHz, making it observable today at low radio frequencies. Redshift allows the 21-cm spectrum to trace the history of the neutral hydrogen gas from the Dark Ages, before the first astrophysical sources, through Cosmic Dawn, when stars, galaxies, and black holes began to form, to the Epoch of Reionization (EoR), at the end of which the 21-cm signal is extinguished.

Recently, the Experiment to Detect the Global EoR Signature (EDGES) reported an absorption trough at frequencies roughly corresponding to the Cosmic Dawn portion of the 21-cm spectrum \citep{Bowman:2018}. The depth and shape of the trough, however, were much different than that predicted by the standard $\Lambda$CDM cosmological model. The unexpected nature of the claimed EDGES detection has led to numerous attempts to explain the result through non-standard physics such as non-gravitational dark matter-baryon scattering \citep{Barkana:2018, Fialkov:2018, Loeb&Munoz:2018, Berlin:2018} or a brighter than expected radio background \citep{Feng&Holder:2019, Ewall-Wice:2018, Fialkov&Barkana:2019, Mebane:2020}. Other studies have found that possible systematic effects \citep{Hills:2018, Bradley:2019, Sims&Pober:2020} or inadequacies in the foreground model used to fit the observations \citep{Tauscher:2020b} could account for the observed trough. Contamination from the leakage of polarized foreground emission could also distort the extracted signal \citep{Spinelli:2019}.

A notoriously challenging aspect of analyzing global 21-cm observations is galactic/extragalactic foreground emission, which is composed primarily of synchrotron radiation and is 4-6 orders of magnitude larger than the 21-cm signal, which has an expected magnitude in the hundreds of mK \citep{Pritchard:2012}. Although the foreground radiation is expected to be intrinsically spectrally smooth due to the power law nature of the synchrotron emission, spatial anisotropies in the emission will be averaged together by wide antenna beams introducing additional spectral structure.

While some have used polynomial-based foreground models to analyze observations (e.g., \citealt{Bowman:2018}), \cite{Tauscher:2018}, hereafter referred to as Paper I, introduces a method to model the foreground through training sets created by varying the characteristics of the beam-weighted foreground model within realistic uncertainties. Singular Value Decompostion (SVD) is then performed on the training set to acquire eigenmodes (hereafter referred to simply as ``modes''), or basis vectors, with which to fit the foreground. A similar approach is used to create a training set and find basis vectors for the 21-cm signal. Observations are then fit with both sets of SVD modes simultaneously in order to extract the 21-cm signal and construct confidence intervals determined by the noise level of the data and the overlap between the models. The code underlying this method, known as \texttt{pylinex},\footnote{\url{https://bitbucket.org/ktausch/pylinex}} is publicly available.

while the SVD approach described in Paper I is sufficiently general to analyze observations from nearly any low frequency global signal experiment, it was developed with the goal of creating a data analysis pipeline for the Dark Ages Polarimeter PathfindER (DAPPER; \citealt{Burns:2020}). DAPPER aims to take advantage of the lunar farside, which is free of ionospheric contamination and terrestrial radio frequency interference \citep{Bassett:2020}, to observe both the Dark Ages and the Cosmic Dawn portions of the 21-cm spectrum. The simulated data presented in this work are constructed to be similar in nature to observations made by DAPPER.

Paper I is the first of a multi-paper series introducing the full pipeline for analyzing 21-cm observations. Additional details of the pipeline concerning constraints on nonlinear signal parameters via a Markov Chain Monte Carlo (MCMC) engine and the benefits of utilizing full Stokes polarization observations and time dependence to decrease the overlap between signal and foreground models are presented in \citealt{Rapetti:2019} (Paper II) and \citealt{Tauscher:2020a} (Paper III), respectively.

While Papers I-III describe how the pipeline is implemented in detail, the analysis rests on the assumption that the training sets have been constructed such that they are representative of the components of a given observation. In this paper we ask how do we determine if the training sets are sufficient to provide an optimal fit to each of the data components?
To answer this question, we describe a novel method for assessing whether or not a collection of training sets is sufficient to precisely extract the 21-cm signal for a given set of observations. Applying this method ensures the robustness of the pipeline by identifying components in the data that are ``outliers'' relative to the training sets and updating the analysis accordingly.

In Section \ref{methodology}, we detail the linear portion of the pipeline for the purpose of outlining our main test to assess the validity of the training sets. We describe how to implement the test when the analysis is performed with or without a prior distribution. In Section \ref{test_cases}, we describe four different cases in which simulated observations are made from different models to disagree with the training sets used for modelling/fitting the data to various degrees, while the results of assessing the disagreement of these cases with our methodology are described in Section \ref{results}. We then summarize and present conclusions in Section \ref{conclusions}.

\section{Methodology}
\label{methodology}

\subsection{Simulations and training sets}
\label{simulations_and_training_sets}

The training sets used by \texttt{pylinex} are constructed for each component of a given data set based on simulations, other observations, and theoretical knowledge. The modes for each component produced by SVD are then used to perform a simultaneous weighted least-square fit to extract the 21-cm signal. The number of modes to use for each component is chosen by minimizing the Deviance Information Criterion (DIC), given by
\begin{equation}
\label{DIC_equation}
    \textup{DIC} = \boldsymbol{\delta}^T\boldsymbol{C}^{-1}\boldsymbol{\delta} + 2N_p,\\
\end{equation}
where $\boldsymbol{\delta} = \boldsymbol{\gamma} - \boldsymbol{y}$ is the residual of the maximum likelihood fit $\boldsymbol{\gamma}$ with respect to the data $\boldsymbol{y}$, $\boldsymbol{C}$ is the noise covariance of the data, and $N_p$ is the number of parameters of the fit (i.e. the total number of SVD modes). Note that if a prior probability distribution is used, we maximize the Bayesian evidence instead of minimizing the DIC (see Section \ref{priors}).

Although the tests outlined in this paper can be applied to any 21-cm experiment analyzed with \texttt{pylinex}, in this paper we focus on simulated observations from a dual-polarization antenna measuring the four stokes parameters I, Q, U, and V from 40 to 120 MHz with 0.5 MHz resolution. While the foreground emission is assumed to be intrinsically unpolarized, the projection of the foreground onto the antenna plane will produce an induced polarization effect. For more details on the advantages of polarization in 21-cm experiments see \cite{Nhan:2019} and \cite{Tauscher:2020a}. For simplicity, we ignore any systematic effects beyond the beam-weighted foreground.\footnote{Paper IV of the series is expected to discuss including an instrument receiver within the training set formalism.} We also ignore any contamination from Earth's ionosphere or terrestrial radio frequency interference (RFI) so that the simulated observations mimic a lunar-based experiment such as DAPPER.

Each simulated observation consists of 96 individual spectra composed of the four Stokes parameters measured at six angles of rotation about four antenna pointing directions.\footnote{The four pointings used are (60$^{\circ}$, 90$^{\circ}$), (60$^{\circ}$, 270$^{\circ}$), (-60$^{\circ}$, 90$^{\circ}$), and (-60$^{\circ}$, 270$^{\circ}$) in galactic coordinates ($b,\ l$).}\textsuperscript{,}\footnote{Since there are 96 spectra each consisting of 161 frequency channels, there are 15,456 data points in each observation.} The data are composed of a beam-weighted foreground (hereafter referred to simply as the ``foreground'') for each of the pointings and rotation angles and the 21-cm signal, which will only appear in Stokes I due to its assumed isotropy. The foregrounds are created from sky maps observed at higher frequencies, which are then interpolated to the relevant band. For further details about the maps see Section \ref{spectral_index}. Gaussian noise following the radiometer equation for 800 total hours of integration divided equally between each pointing and rotation angle is added to the foreground and 21-cm components to complete the simulated observation. For a derivation of the noise level on each of the four Stokes parameters, see Appendix C of \cite{Tauscher:2020a}.

Training sets are created in an analogous way to the simulated observations, but are separated by component (i.e. foreground and 21-cm signal) and do not include noise. For the foregrounds, we can choose to vary the characteristics of the antenna beam or the spectral index model used to perform the interpolation in frequency-space. Although in principle the base sky map could also be varied, we do not investigate that aspect in this work. Further details regarding the foreground models are again given in Section \ref{spectral_index}.

\newpage
\subsection{Statistical goodness-of-fit measures}

In this section we present various goodness-of-fit measures that can used to assess fits performed by the pipeline.

\subsubsection{Signal bias statistic}

\renewcommand{\baselinestretch}{0.5}
\begin{figure*}
    \centering
    \includegraphics[width=\textwidth]{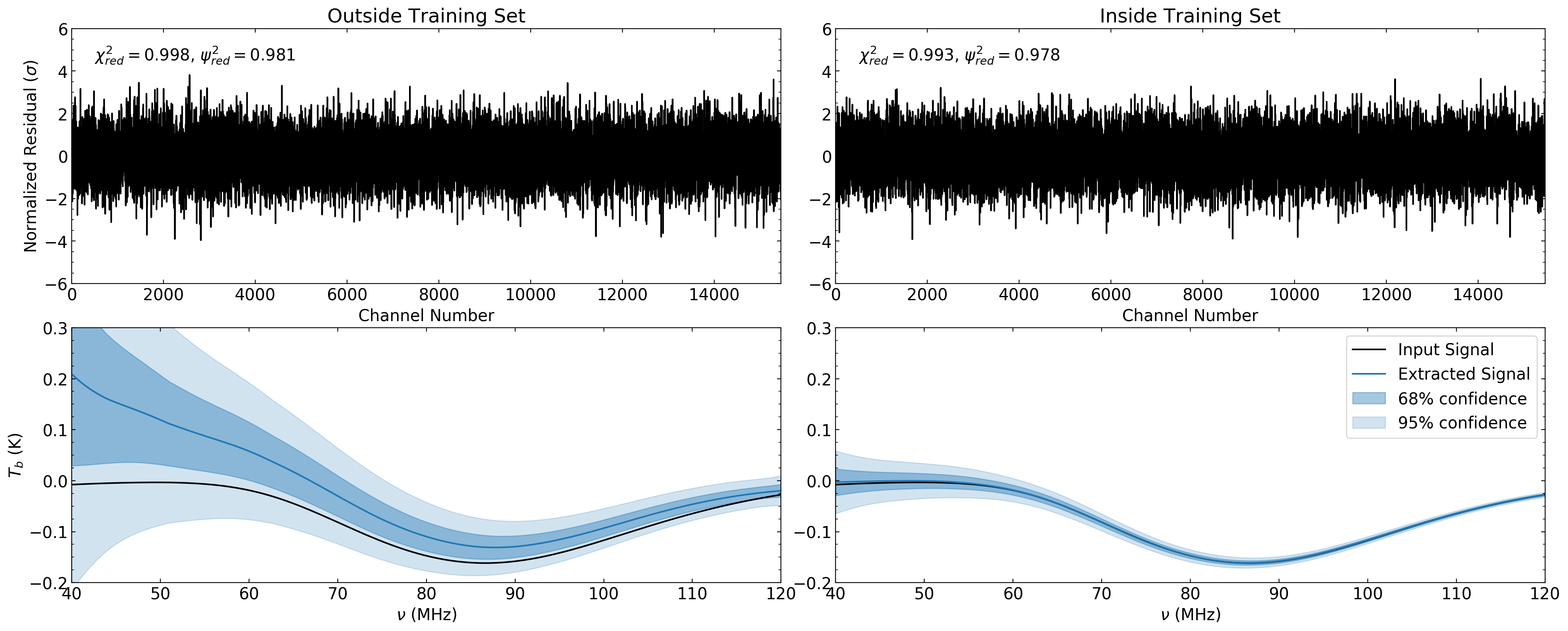}
    \caption{\textit{Left}: Example of a fit when the foreground used to create the simulated data does not match the foreground training set. The top panel shows the residual of the fit that minimizes the DIC. The residual is normalized by the noise level, which varies across the band following the radiometer equation for a dual-dipole experiment. The values of $\chi^2_{\textup{red}}$ and $\psi^2_{\textup{red}}$ are indicated. In the bottom panel the reconstructed 21-cm signal with 68\% and 95\% confidence intervals is compared to the input signal. Note that in the top panel ``channel number'' refers to the concatenation of data points over pointings, rotations, Stokes polarizations, and frequencies. \textit{Right}: Example of a fit when the foreground used to create the simulated data is described well by the foreground training set. Again, the top panel shows the residual of the fit while the bottom panel shows the extraction of the 21-cm signal component. The comparison of the left and right examples shows that while the 21-cm signal is extracted accurately in both cases, the precision is greatly increased when the input data matches the training set.}
    \label{fig:failure_example}
\end{figure*}

In order to directly quantify the success of the algorithm in extracting the 21-cm signal from the data, we follow Paper I in defining the signal bias statistic, $\varepsilon$, as
\begin{equation}
    \varepsilon = \sqrt{\frac{1}{n_{\nu}}\sum_{i=1}^{n_{\nu}}\frac{[(\boldsymbol{\gamma}_{\textup{21-cm}} - \boldsymbol{y}_{\textup{21-cm}})_i]^2}{(\boldsymbol{\Delta}_{\textup{21-cm}})_{ii}}}
\end{equation}
where $n_{\nu}$ is the number of frequencies at which the signal is measured, $\boldsymbol{y}_{\textup{21-cm}}$ is the true (input) signal, and $\boldsymbol{\gamma}_{\textup{21-cm}}$ and $\boldsymbol{\Delta}_{\textup{21-cm}}$ are the maximum likelihood reconstruction and covariance of the signal in frequency space. A signal bias statistic of $\varepsilon$ corresponds approximately to a bias level of $\varepsilon \sigma$ between the reconstructed and the true signal.

We use the signal bias statistic to calculate the $\sigma$-level necessary to reach a given percentage probability for confidence intervals. In order to do this, we create the cumulative distribution function (CDF) of the signal bias statistic for 5000 data realizations made from curves taken from the training sets. We then find the value of the signal bias statistic where the CDF crosses $p/100$ and multiply the diagonal elements of $\boldsymbol{\Delta}_{\textup{21-cm}}$ by this factor to produce the $p$\% confidence band.

\subsubsection{Chi-Squared}
\label{chi_squared}

The commonly used goodness-of-fit statistic $\chi^2_{\textup{red}}$ can be defined as
\begin{equation}
\label{chi_squared_equation}
    \chi_{\textup{red}}^2 = \frac{\boldsymbol{\delta}^T\boldsymbol{C}^{-1}\boldsymbol{\delta}}{N_c - N_p},
\end{equation}
where $N_c$ is the number of data channels while $N_p$, $\boldsymbol{C}$, and $\boldsymbol{\delta}$ are defined as in Equation \ref{DIC_equation}. Rather than assessing the accuracy of the signal reconstruction like the signal bias statistic, $\chi^2_{\textup{red}}$ measures the overall quality of the simultaneous fit to both components.

\subsubsection{Psi-squared}
\label{psi_squared}

Although the $\chi^2_{\textup{red}}$ statistic described in Section \ref{chi_squared} is nearly ubiquitous in model-fitting, it may not recognize unmodeled wideband features in the data close to the noise level. Whereas $\chi^2$ measures only the magnitude of the residuals (in the case that $\boldsymbol{C}$ is diagonal), the reduced $\psi^2$ statistic, introduced in \cite{Tauscher:2018b}, probes channel-to-channel correlations and can more easily detect residual features close to the noise level.

The $\psi^2_{\textup{red}}$ statistic is defined through the correlation vector $\boldsymbol{\rho}$, whose $q^{\textup{th}}$ component is given by
\begin{equation}
    \rho_q = \frac{1}{N_c - q}\sum_{k=1}^{N_c - q}\Delta_k\Delta_{k+q},
\end{equation}
where, here, $\mathbf{\Delta} = \mathbf{C}^{-1/2}\boldsymbol{\delta}$ and $\Delta_k$ is the $k^{\textup{th}}$ element of $\boldsymbol{\Delta}$. The $\psi^2_{\textup{red}}$ statistic can then be defined as
\begin{equation}
    \psi_{\textup{red}}^2 = \frac{1}{N_c - 1}\sum_{q=1}^{N_c - 1}\Big(\frac{\rho_q}{\sigma_{\rho_q}}\Big)^2, 
\end{equation}
where $\sigma_{\rho_q} = \sqrt{\textup{Var}[\rho_q]}$.\footnote{A step-by-step procedure for calculating $\psi_{\textup{red}}^2$ is outlined in Appendix A of \cite{Tauscher:2018b}.} In the limit of large $N_c$, the expectation value and variance of $\psi^2_{\textup{red}}$ are $\textup{E}[\psi_{\textup{red}}^2] = 1$ and $\textup{Var}[\psi_{\textup{red}}^2] = 14/N_c$.

\subsubsection{Limitations}

Each of the statistics discussed above have limitations in their usefulness. The signal bias statistic requires knowledge of the true form of the signal, which is not available when fitting experimental measurements. Alternatively, while $\chi^2_{\textup{red}}$ and $\psi^2_{\textup{red}}$ do not require any prior knowledge of the underlying components, they only measure the fit to the full data.

The examples shown in Figure \ref{fig:failure_example} illustrate the limitations of $\chi^2_{\textup{red}}$ and $\psi^2_{\textup{red}}$ in more detail. Fits to two different simulated observations are shown. One observation uses a foreground realization that lies outside of the training set, while the other uses a foreground model within the training set.\footnote{More specifically, the outside of training set example uses the green beam shown in the top panel of Figure \ref{fig:beams}, while the inside training set example uses the orange beam.} The 21-cm signal used for both cases is drawn directly from the signal training set (for more details on how the signal training set is generated, see Section \ref{signals}). As shown in the top panels, the $\chi^2_{\textup{red}}$ and $\psi^2_{\textup{red}}$ values for the residuals of the fits are consistent with Gaussian noise in both cases. However, the constraints on the 21-cm signal vary significantly between the two examples. In each of the fits, the signal is extracted accurately (in the sense that the bias between the reconstruction and the input is consistent with the uncertainty), but when the foreground is outside of the training set, the signal reconstruction is much less precise (the uncertainty bands are much wider). We see from these examples that $\chi^2_{\textup{red}}$ and $\psi^2_{\textup{red}}$ are insufficient to detect when the signal extraction is suboptimal.

\subsubsection{Numbers of SVD modes}

The difference in precision between the two examples shown in Figure \ref{fig:failure_example} can be explained by the way in which the pipeline chooses the number of SVD modes to fit each component. As outlined in Section \ref{simulations_and_training_sets}, the number of SVD modes to use for the fit is chosen by minimizing the DIC over a grid with axes that correspond to the individual components of the data. Since our simulated data contain only foreground and signal components, this will be a two dimensional grid. By fitting data realizations made using various training set curves with the training set's SVD modes, we can create a distribution of the numbers of modes required to fit the training sets. If the training sets match the data well, then the numbers of SVD modes required to fit the data should fall somewhere within this distribution. However, if the training sets do not represent the data well, such as in the example in the left panels of Figure \ref{fig:failure_example}, then additional modes will be required to fit the mismatch between the training set and the data. While the full data may still be fit well with this larger number of modes, the expense is that the uncertainties in the reconstruction of the signal are increased.

It is important to note that the pipeline has not necessarily ``failed'' when more modes are required to fit the data. While the large uncertainties on the signal reconstruction are not ideal, the reconstruction is still accurate in the sense that the true signal lies within a reasonable uncertainty on the reconstruction. True failure of the pipeline (i.e. when the signal extraction is significantly biased with spuriously small uncertainties) occurs when the number of modes for the pipeline to search is limited to a smaller number than is required to fully describe the data. In this case the DIC will likely choose a number of modes on the edge of the allowed grid and return an inaccurate signal reconstruction. For this reason, caution should be exercised whenever the chosen number of modes is found to be on the edge of the allowed grid, in which case the grid likely needs to be expanded. The connection between the numbers of SVD modes chosen for a given fit and the precision of the signal reconstruction leads us to investigate the numbers of SVD modes as a criterion for assessing the quality of fits performed by the pipeline.

\subsection{Priors}
\label{priors}

In some cases, it may be desirable to include prior information when performing a fit. Due to the fact that the fit is a weighted combination of SVD modes, the priors must constrain the coefficients of the SVD modes rather than the parameters of the underlying model. A useful method for creating a prior distribution on the coefficients is to derive the prior from the training set for each component. In order to produce the prior, each curve in the training set is fit with the SVD modes from its respective training set. The mean $\boldsymbol{\mu}$ and covariance $\boldsymbol{\Lambda}$ of the coefficients are then used to define the multivariate Gaussian prior distribution.\footnote{Due to numerical issues associated with inverting the full covariance matrix $\boldsymbol{\Lambda}$, which is necessary to apply the prior distribution (see Equation \ref{prior_equation}), we use only the variances of the parameters (i.e. the diagonal elements of $\boldsymbol{\Lambda}$).} For the exact, analytical forms of the mean and variance of the SVD mode coefficients, see Appendix C of \cite{Tauscher:2018}.

Without priors, the relative magnitudes of each component are able to vary freely. While fits performed without priors can produce biased results if the training sets cannot sufficiently describe the shapes of the components, the use of a prior derived from the training set applies constraints on both the shapes and the magnitudes of the components. Thus, caution should be taken to ensure that there is a high level of confidence that the given training set is sufficient to describe both the shape and magnitude of the component before a training set prior is applied.\footnote{In principle, if one is uncertain that the training set is fully representative of the data, it may be possible to apply a ``weak'' prior, which falls in between the highly constrained and uniform priors used in the examples here. A complicating factor in applying this method is that the posterior will no longer be Gaussian if the prior takes a non-Gaussian form.}

Note that when using a prior distribution, instead of minimizing the DIC in order to choose the number of SVD modes, we instead maximize the Bayesian evidence $\mathcal{Z}$, given by
\begin{subequations}
\label{prior_equation}
\begin{align}
    \mathcal{Z} &= \int \mathcal{L}(\boldsymbol{x})\pi(\boldsymbol{x})\textup{d}^{N_p}\boldsymbol{x},\\
    \mathcal{L}(\boldsymbol{x}) &\propto \textup{exp}
    \Big\{-\frac{1}{2}[\boldsymbol{y} - \mathcal{M}(\boldsymbol{x})]^T\boldsymbol{C}^{-1}[\boldsymbol{y} - \mathcal{M}(\boldsymbol{x})]\Big\},\\
    \pi(\boldsymbol{x}) &\propto \textup{exp}
    \Big\{-\frac{1}{2}[\boldsymbol{x} - \boldsymbol{\mu}]^T\boldsymbol{\Lambda}^{-1}[\boldsymbol{x} - \boldsymbol{\mu}]\Big\},
\end{align}
\end{subequations}
where the parameters $\boldsymbol{x}$ are the coefficients of the SVD modes, $\mathcal{L}(\boldsymbol{x})$ is the likelihood,  $\mathcal{M}(\boldsymbol{x})$ is the model composed of the SVD modes, and $\pi(\boldsymbol{x})$ is the prior distribution. Note that the dimensionality $N_p$ of the parameter space will change when the number of SVD modes to be fit is varied. Maximizing the evidence ensures that the prior is properly taken into account. When using a training set prior, variations in modes unimportant to the training set are suppressed, but a larger number of modes may still be chosen by the evidence. Thus, the number of SVD modes may no longer provide as much information about the quality of the estimate of the signal. However, if the training sets do not encompass the true form of one or more of the components, the constraints provided by the prior will prevent the full data from being fit well. In this case, goodness-of-fit statistics such as $\chi^2_{\textup{red}}$ or $\psi^2_{\textup{red}}$ will be able to identify a poor fit. Further details regarding the use of these metrics for fits with priors are given in Section \ref{test_cases}.

\tikzstyle{block} = [rectangle, rounded corners, minimum width=3cm, minimum height=1cm, text centered, draw=black]
\tikzstyle{arrow} = [thick,->,>=stealth]
\begin{figure}
    \centering
    \resizebox{\columnwidth}{!}{
    \begin{tikzpicture}[node distance=2cm]
      \node (step1) [block, align=center] {Create/alter training sets\\for all components of the data};
        \node (priors) [block, below of=step1, align=center] {Are priors being used\\for any of the components?};
        \node (step2a) [block, below of=priors, align=center, xshift=-2.5cm] {Create reference distribution\\of numbers of SVD modes};
        \node (step2b) [block, below of=priors, align=center, xshift=2.5cm] {Create reference distribution\\of $\chi^2$ and/or $\psi^2$};
        \node (step3) [block, below of=step2a, align=center, xshift=2.5cm] {Fit data of interest and\\calculate p-value};
        \node (step4) [block, below of=step3, align=center] {Is the fit consistent with the reference\\distribution to the desired significance level?};
        \node (step5) [block, below of=step4, align=center] {Accept fit and signal extraction};
        \draw [arrow] (step1) -- (priors);
        \draw [arrow] (priors) -- node[above=1pt] {no} (step2a);
        \draw [arrow] (priors) -- node[above=1pt] {yes} (step2b);
        \draw [arrow] (step2a) -- (step3);
        \draw [arrow] (step2b) -- (step3);
        \draw [arrow] (step3) -- (step4);
        \draw [arrow] (step4) -- node[anchor=west] {yes} (step5);
        \draw [arrow] (step4) -- node[above=2pt] {no} ++(5.5,0) |- (step1);
    \end{tikzpicture}
    }
    \caption{A flow chart depicting an outline of the suggested steps for the tests outlined in this paper.}
    \label{fig:flow_chart}
\end{figure}
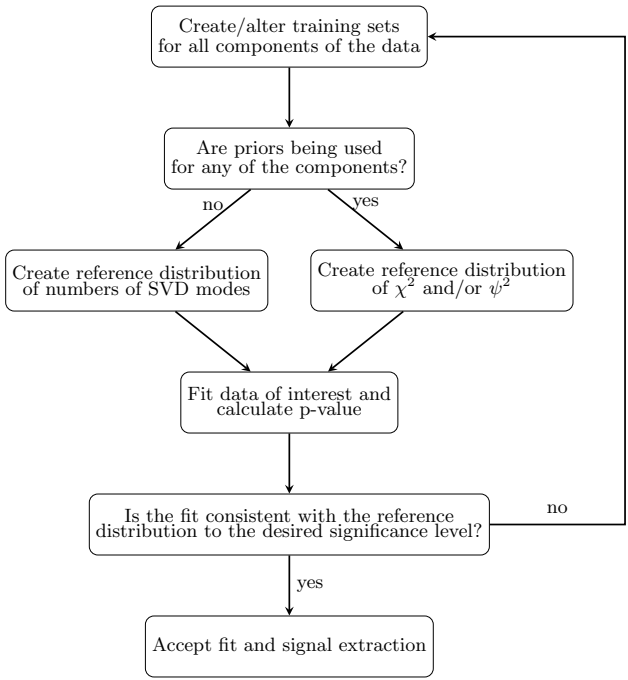

\subsection{Test strategy}
\label{test_outline}

In this section, we present a step-by-step outline for assessing the quality of the signal extraction in an observational setting when the true form of the signal is unknown. This test can be performed for fits with or without priors, however we suggest that the test be performed without a prior distribution first due to the fact that the use of an incorrect prior distribution will lead to significant bias in the extracted signal, as described in Section \ref{priors}. If the fit ``passes'' the test as described below, one may wish to include a training set prior to possibly improve the quality of the signal reconstruction further.
\begin{enumerate}
    \item Fit many data realizations created from the training set curves. If no prior is used, create a distribution of the chosen numbers of SVD modes. This distribution will have the same dimensionality as the number of components in the data. If a prior distribution is used, create a distribution of a given goodness-of-fit metric for the full data such as $\chi^2_{\textup{red}}$ or $\psi^2_{\textup{red}}$.
    \item Fit the data of interest and determine the p-value. The p-value is calculated as the portion of the distribution of the numbers of SVD modes chosen or the $\chi^2_{\textup{red}}$ or $\psi^2_{\textup{red}}$ statistics from step 1 more extreme than the given fit.
    \begin{enumerate}
        \item If the p-value is greater than the desired significance level, accept the fit.
        \item If the p-value is less than the desired significance level, the signal extraction may be improved by updating the training sets.
    \end{enumerate}
    \item If needed, alter the training sets accordingly and return to step 1.
\end{enumerate}
A flow chart of the steps is presented in Figure \ref{fig:flow_chart}. As outlined, the tests described above essentially constitute a null hypothesis significance test. In this case the null hypothesis is that the training sets are sufficient to fit the data either with or without priors. Thus, the p-value can be calculated from the portion of the reference distribution outside of the given fit, which is interpreted as the probability of obtaining a result at least as extreme under the null hypothesis. While the significance level to reject the null hypothesis may be chosen arbitrarily, we adopt 0.05 as a fiducial value. A smaller value for significance level will likely prevent fits from being rejected when the data lies near the edge of the training sets, but may be more susceptible to cases that should be rejected but happen to lie near the edge of the reference distribution. The examples in Section \ref{test_cases} illustrate the test in practice.

\section{Test Cases}
\label{test_cases}

The following subsections will each present examples in which one or more components of the input data differs from its respective training set to varying degrees. In general, the orange examples (see Figures \ref{fig:beams}, \ref{fig:spectral_index}, \ref{fig:signals}, \ref{fig:beam_and_signal}) are chosen to be fiducial models drawn directly from the training sets, while the red, purple, and green examples lie outside of the training sets. Although we generally refer to the beam-weighted sky as the foreground component, the beam and the sky are modeled separately and then combined to make the training set. For the purposes of this paper, we will investigate the beam and the sky model as separate examples. The implications of the examples described below are discussed in Section \ref{results}.

\subsection{Beams}
\label{beams}

\begin{figure*}
    \centering
    \includegraphics[width=0.32\textwidth]{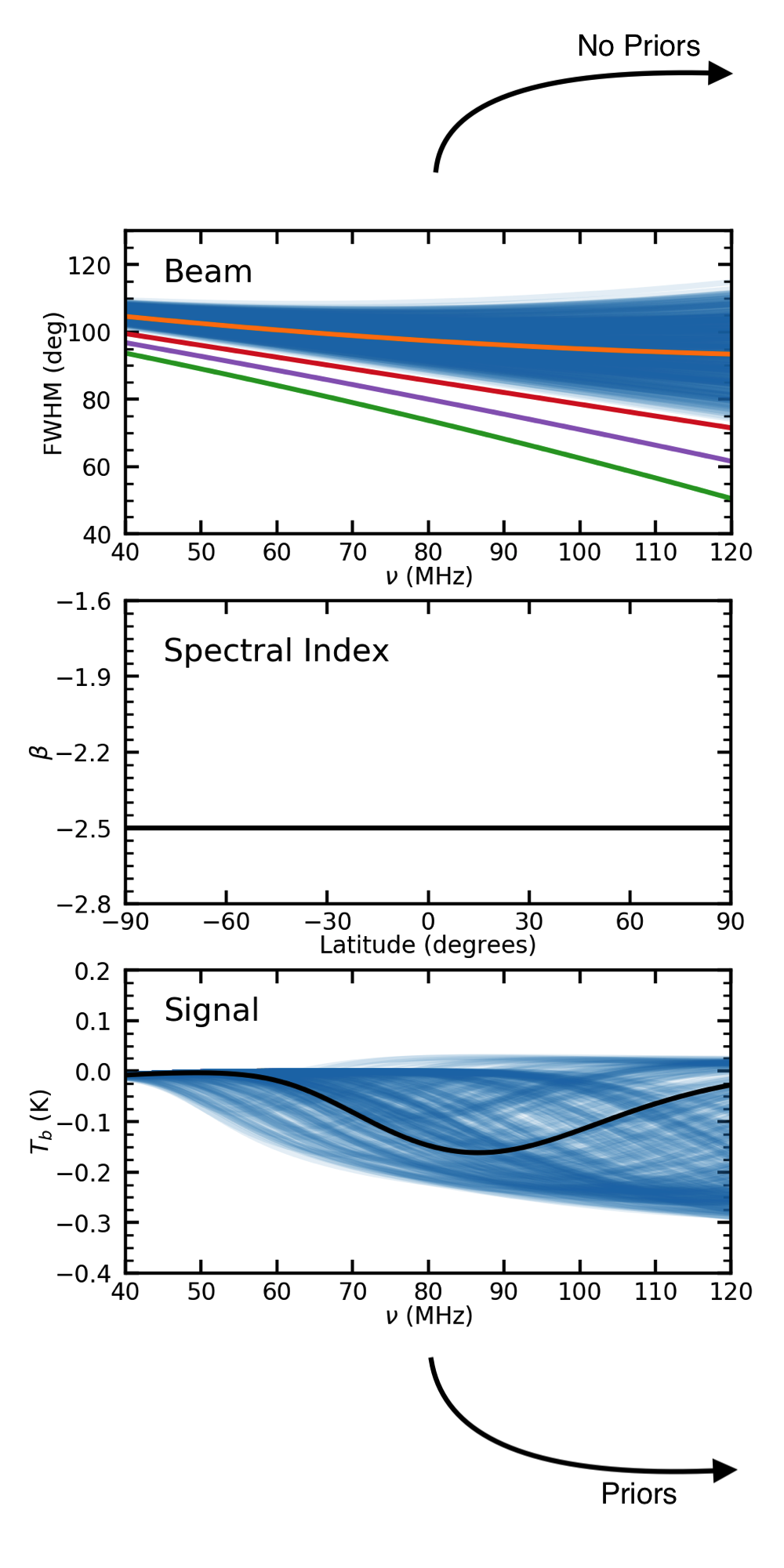}
    \includegraphics[width=0.66\textwidth]{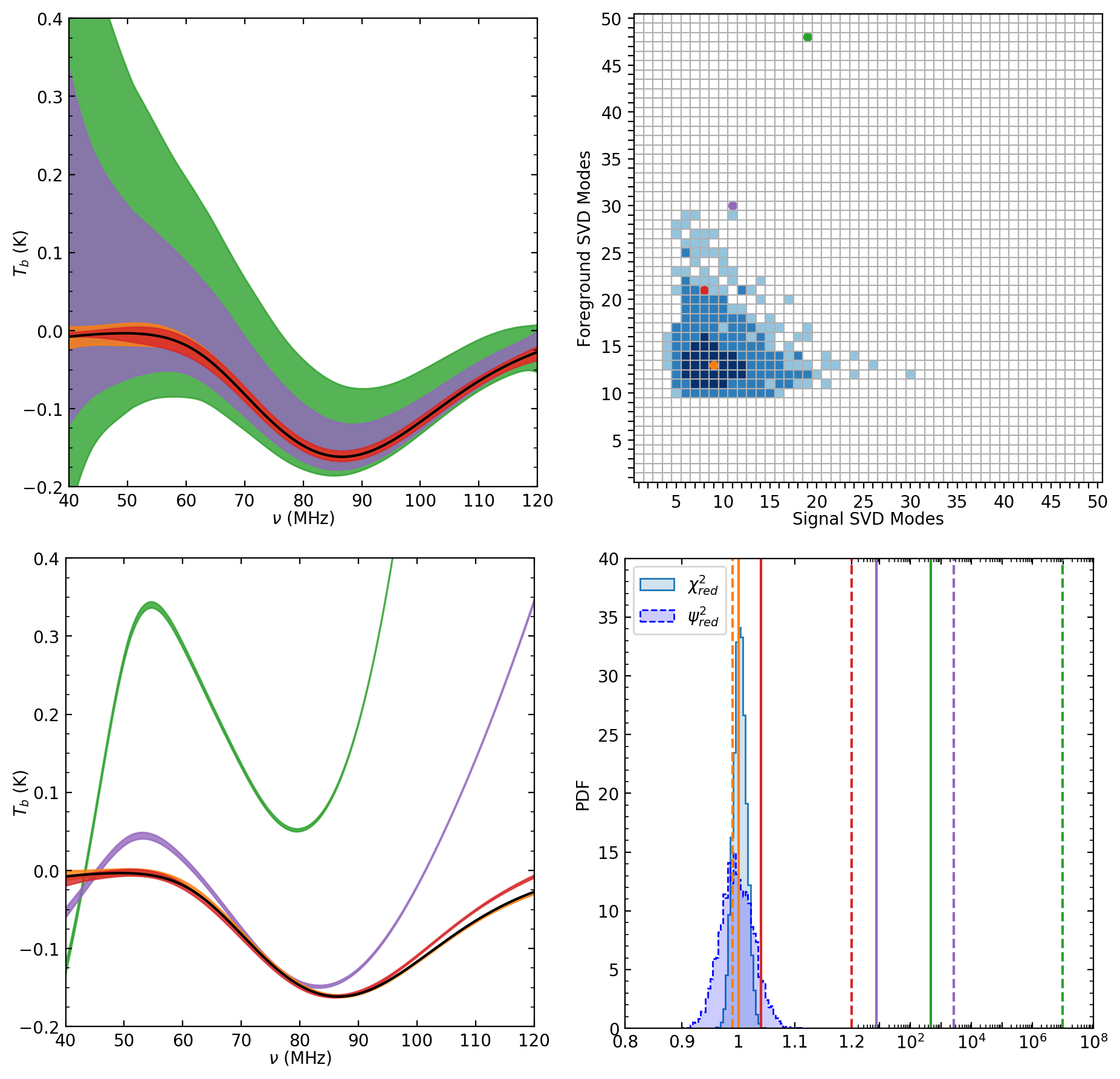}
    \caption{\textit{Left}: Input components used to create the simulated data. The orange, red, purple, and green beams along with the black spectral index and signal are used to create four realizations to be fit using the SVD modes from the training sets shown in blue. \textit{Middle}: 95\% confidence signal reconstructions produced by the pipeline for the four simulated data realizations both without (top) and with (bottom) training set priors. The color corresponds to the beam in the left panel used to create the realization. \textit{Right}: Results of the tests without (top) and with (bottom) priors outlined in Section \ref{test_outline}. The top panel shows the SVD mode distribution when data realizations are made using training set curves (blue contours) as well as the numbers of modes chosen for the fits using the four example beams (colored points). The blue contours enclose 68\%, 95\%, and 99.7\% of the 5000 fits performed. The bottom panel shows values of $\chi^2_{\textup{red}}$ (solid lines) and $\psi^2_{\textup{red}}$ (dashed lines) for fits including priors using the four example beams compared to distributions of each statistic from 5000 data realizations made using training set curves. The x-axis scale is linear from $0.8 - 1.2$ and logarithmic above 1.2.}
    \label{fig:beams}
\end{figure*}

For the first test, we keep the sky map and spectral index constant while altering the antenna beam. We use the Haslam all-sky survey performed at 408 MHz \citep{Haslam:1982} with a constant spectral index of -2.5. For the sake of simplicity, we use Gaussian beams, which can be defined entirely at a given frequency by their full width at half maximum (FWHM). The spatial pattern of a Gaussian beam in terms of the FWHM is given by
\begin{equation}
\label{gaussian_beam_equation}
    B_{\textup{Gaussian}} \propto \exp \Bigg\{-\ln 2 \bigg[\frac{\theta}{\textup{FWHM}(\nu)/2}\bigg]^2\Bigg\},
\end{equation}
where $\theta$ is the angle away from the pointing direction. The top left panel of Figure \ref{fig:beams} shows all of the beams used to create the beam-weighted foreground training set in blue, while four beams to be tested are shown in orange, red, purple, and green. The FWHM of all of the beams follow a quadratic function in frequency:
\begin{equation}
    \textup{FWHM}(\nu) = a_0 + a_1 \nu + a_2 \nu^2,
\end{equation}
where the coefficients $a_0$, $a_1$, and $a_2$ are varied to create the different beams. The beam training set was created by drawing random values from the following distributions: $a_0 \sim$ unif(115, 116.5), $a_1 \sim$ unif(-0.35, -0.25), $a_2 \sim$ unif(0, $2 \times 10^{-3}$). The input signal and signal training set, which were created using the Accelerated Reionization Era Simulations (\texttt{ares}; \citealt{Mirocha:2014}) code\footnote{\url{https://github.com/mirochaj/ares}}, are also shown in Figure \ref{fig:beams}. More details on 21-cm signal models can be found in Section \ref{signals}.

First, we perform the fits without using a prior distribution. The 95\% confidence reconstruction of the signal for each of the four data realizations made using the colored beams is shown in the upper middle panel of Figure \ref{fig:beams}. Next, we create the distribution of the numbers of foreground and signal SVD modes used to fit simulated data made from training set curves (i.e. step 1 of the test outlined in Section \ref{test_outline}), which is shown in the upper right panel of Figure \ref{fig:beams}. The blue contours indicate the training set distribution, while the locations of fits to data created using the orange, red, purple, and green beams are shown in their respective colors.

Next, we repeat the analysis with the exact same data realizations, but apply priors derived from the training sets, as described in Section \ref{priors}. The signal reconstructions for these fits are shown in the lower middle panel of Figure \ref{fig:beams}. The lower right panel exhibits the distributions for both $\chi^2_{\textup{red}}$ and $\psi^2_{\textup{red}}$ created using the training set data realizations\footnote{We assume that the values of the statistics for the training set realizations follow the underlying distributions for $\chi^2_{\textup{red}}$ and $\psi^2_{\textup{red}}$.} as well as the values of these statistics for each of the four examples.

\subsection{Spectral Indices}
\label{spectral_index}

\begin{figure*}
    \centering
    \includegraphics[width=0.32\textwidth]{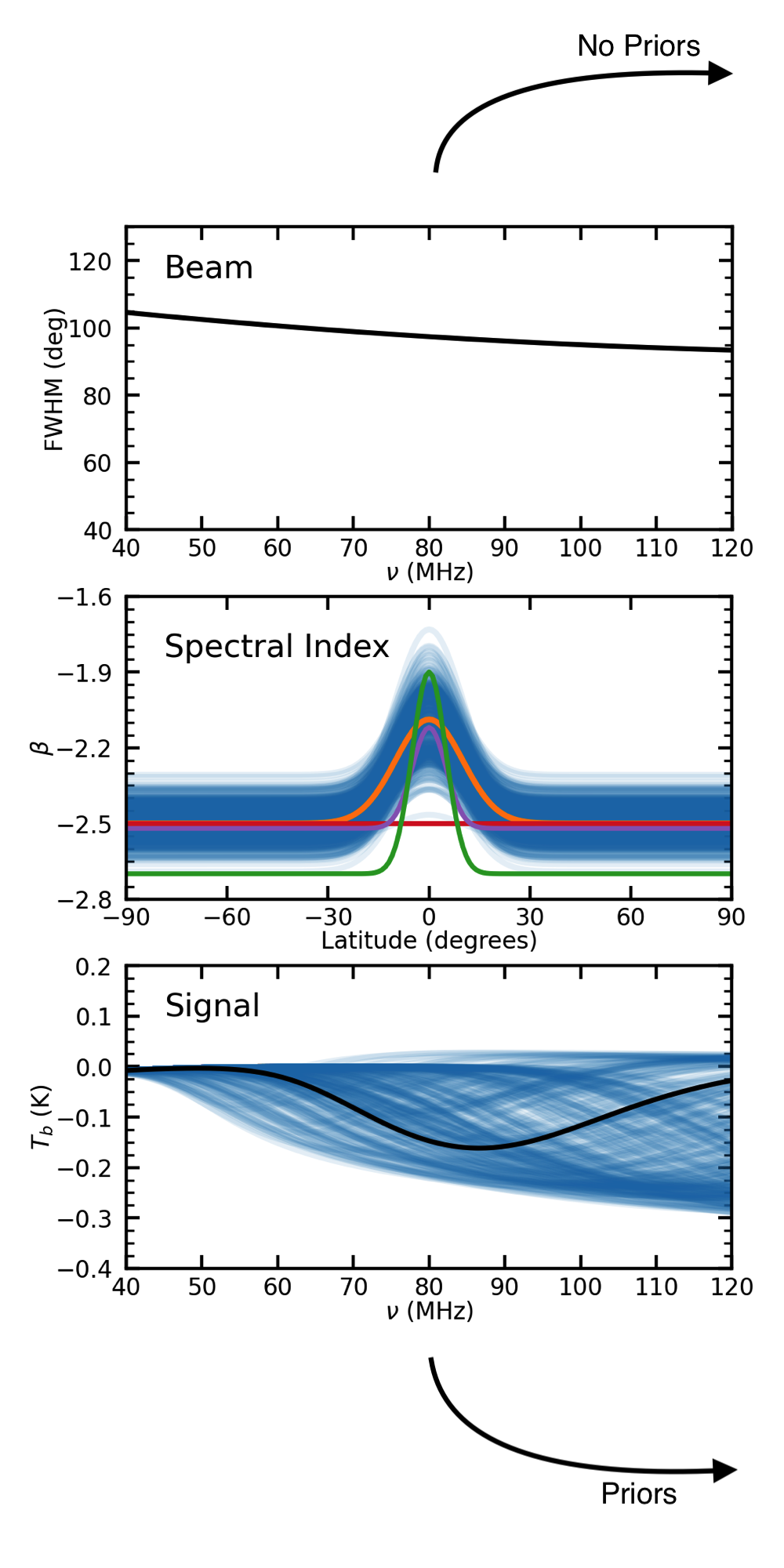}
    \includegraphics[width=0.66\textwidth]{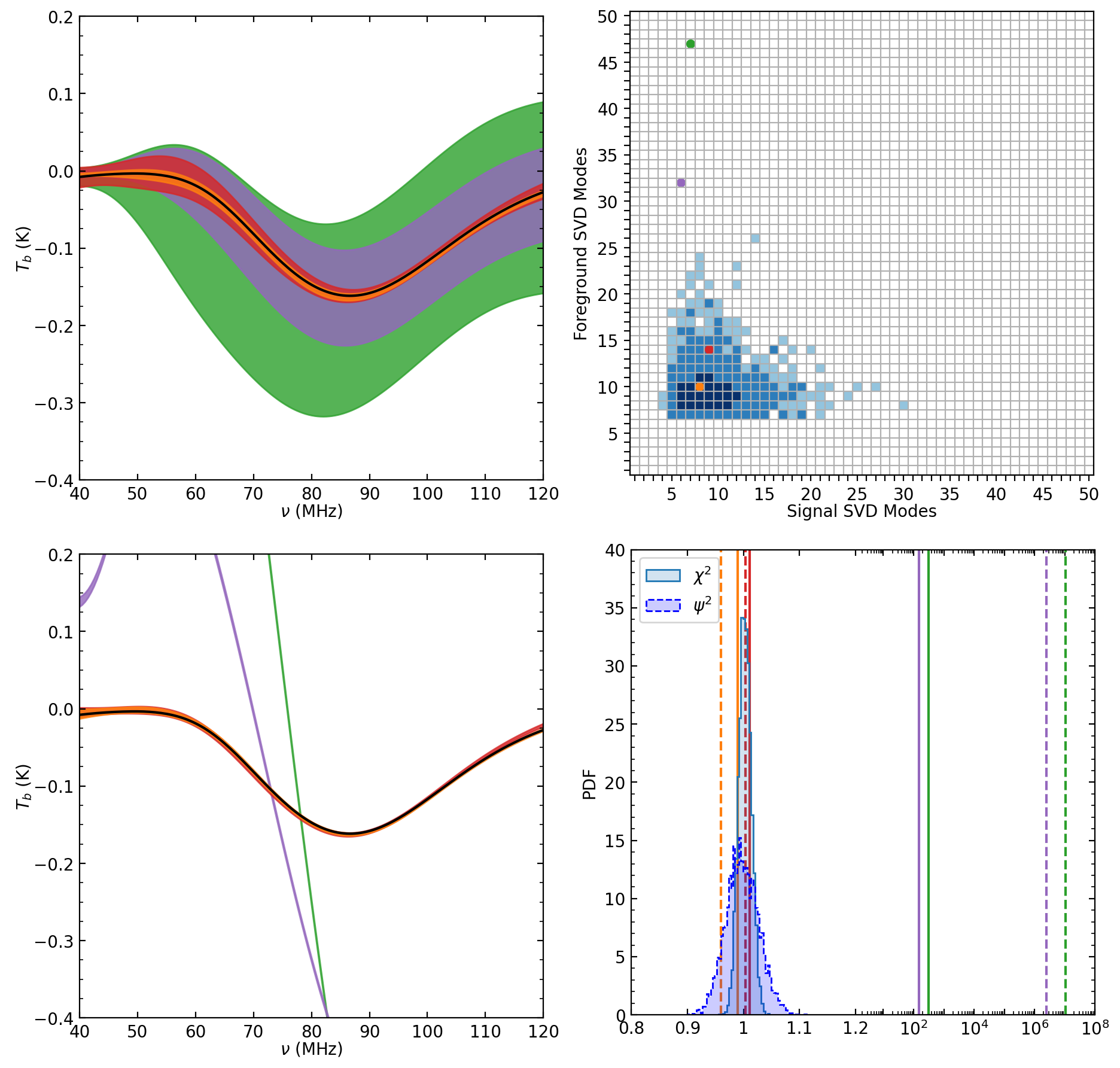}
    \caption{\textit{Left}: Input components used to create the simulated data. The orange, red, and purple spectral index models along with the black beam and signal are used to create three realizations to be fit using the SVD modes from the training sets shown in blue. \textit{Middle}: 95\% confidence signal reconstructions produced by the pipeline for the three simulated data realizations both without (top) and with (bottom) training set priors. The color corresponds to the spectral index model in the left panel used to create the realization. \textit{Right}: Results of the tests without (top) and with (bottom) priors outlined in Section \ref{test_outline}. The top panel shows the SVD mode distribution when data realizations are made using training set curves (blue contours) as well as the numbers of modes chosen for the fits using the three example spectral index models (colored points). The blue contours enclose 68\%, 95\%, and 99.7\% of the 5000 fits performed. The bottom panel shows values of $\chi^2_{\textup{red}}$ (solid lines) and $\psi^2_{\textup{red}}$ (dashed lines) for fits including priors using the three example spectral index models compared to distributions of each statistic from 5000 data realizations made using training set curves. The x-axis scale is linear from $0.8 - 1.2$ and logarithmic above 1.2.}
    \label{fig:spectral_index}
\end{figure*}

\begin{figure*}
    \centering
    \includegraphics[width=0.32\textwidth]{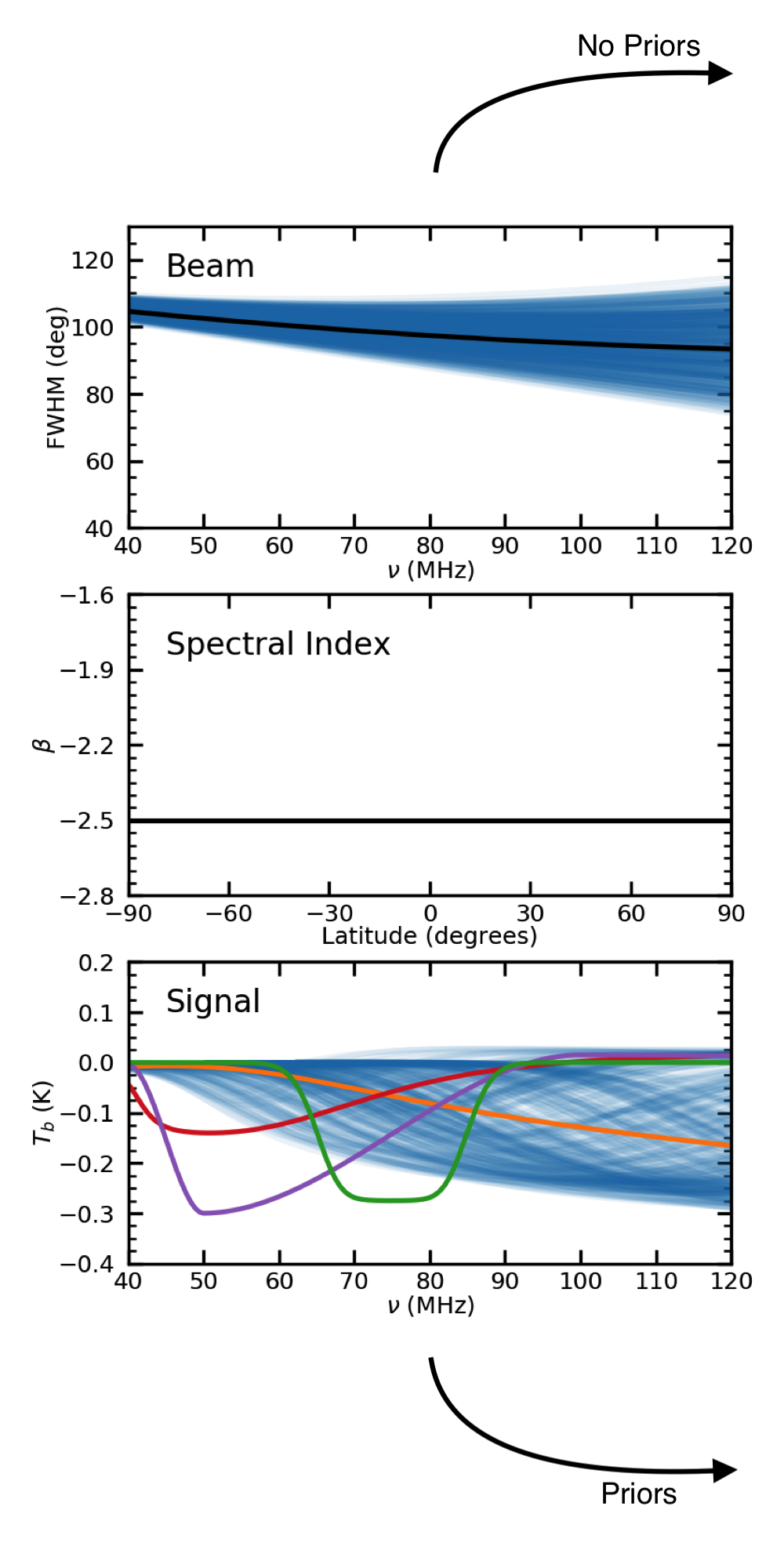}
    \includegraphics[width=0.66\textwidth]{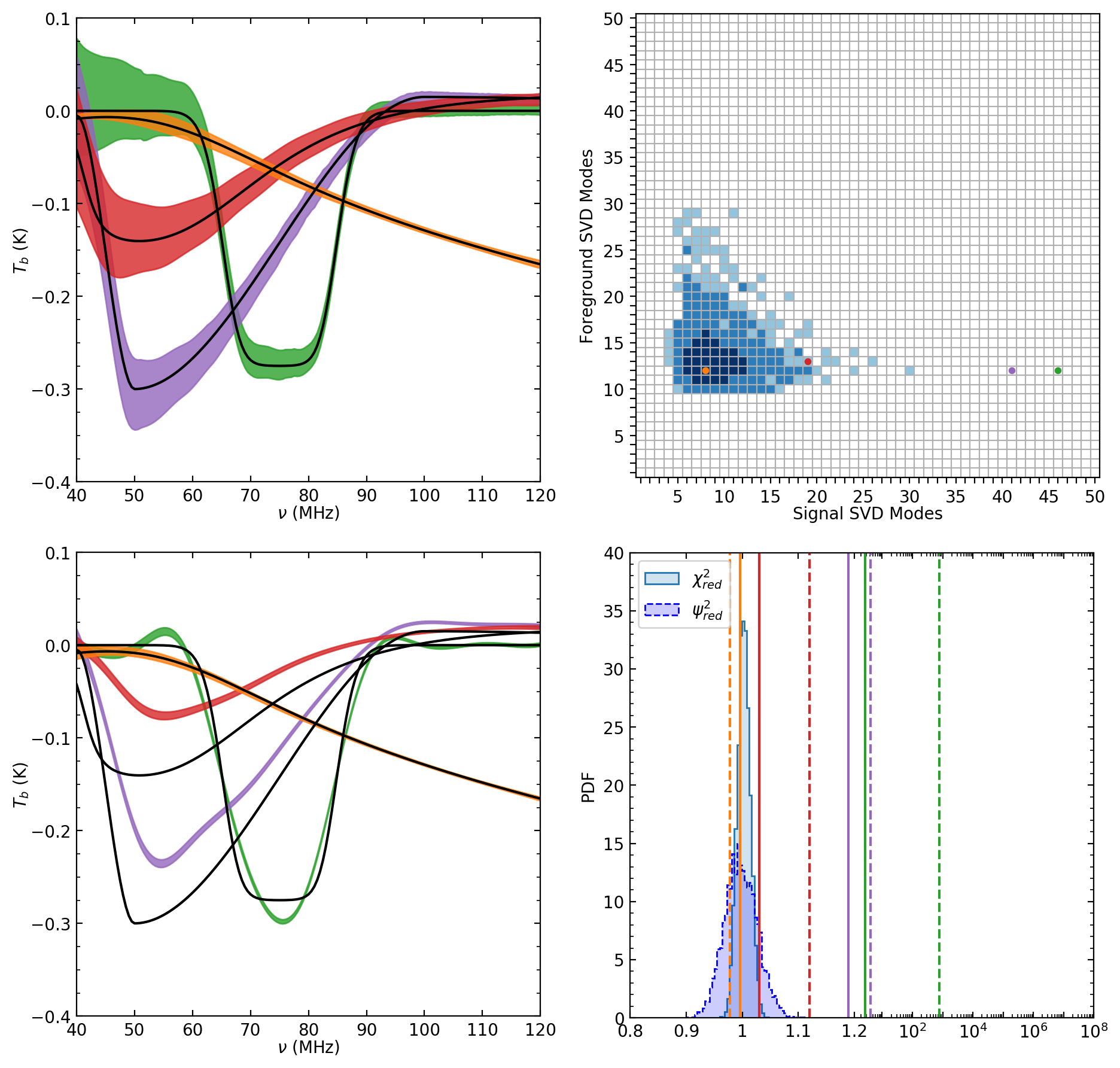}
    \caption{\textit{Left}: Input components used to create the simulated data. The orange, red, purple, and green signals along with the black beam and spectral index are used to create four realizations to be fit using the SVD modes from the training sets shown in blue. \textit{Middle}: 95\% confidence signal reconstructions produced by the pipeline for the four simulated data realizations both without (top) and with (bottom) training set priors. The color corresponds to the signal in the left panel used to create the realization. \textit{Right}: Results of the tests without (top) and with (bottom) priors outlined in Section \ref{test_outline}. The top panel shows the SVD mode distribution when data realizations are made using training set curves (blue contours) as well as the numbers of modes chosen for the fits using the four example signals (colored points). The blue contours enclose 68\%, 95\%, and 99.7\% of the 5000 fits performed. The bottom panel shows values of $\chi^2_{\textup{red}}$ (solid lines) and $\psi^2_{\textup{red}}$ (dashed lines) for fits including priors using the four example signals compared to distributions of each statistic from 5000 data realizations made using training set curves. The x-axis scale is linear from $0.8 - 1.2$ and logarithmic above 1.2.}
    \label{fig:signals}
\end{figure*}

\begin{figure*}
    \centering
    \includegraphics[width=0.32\textwidth]{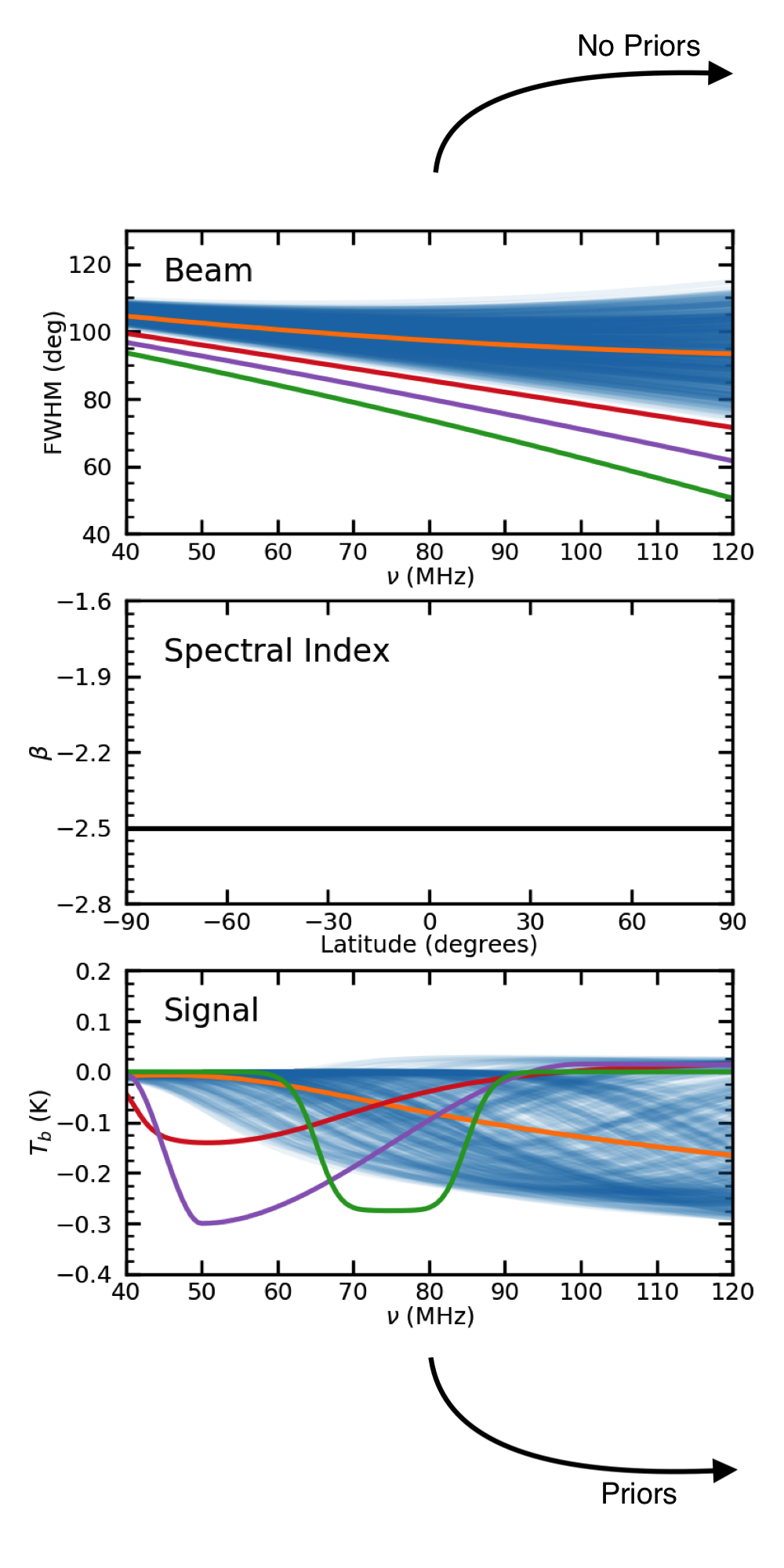}
    \includegraphics[width=0.66\textwidth]{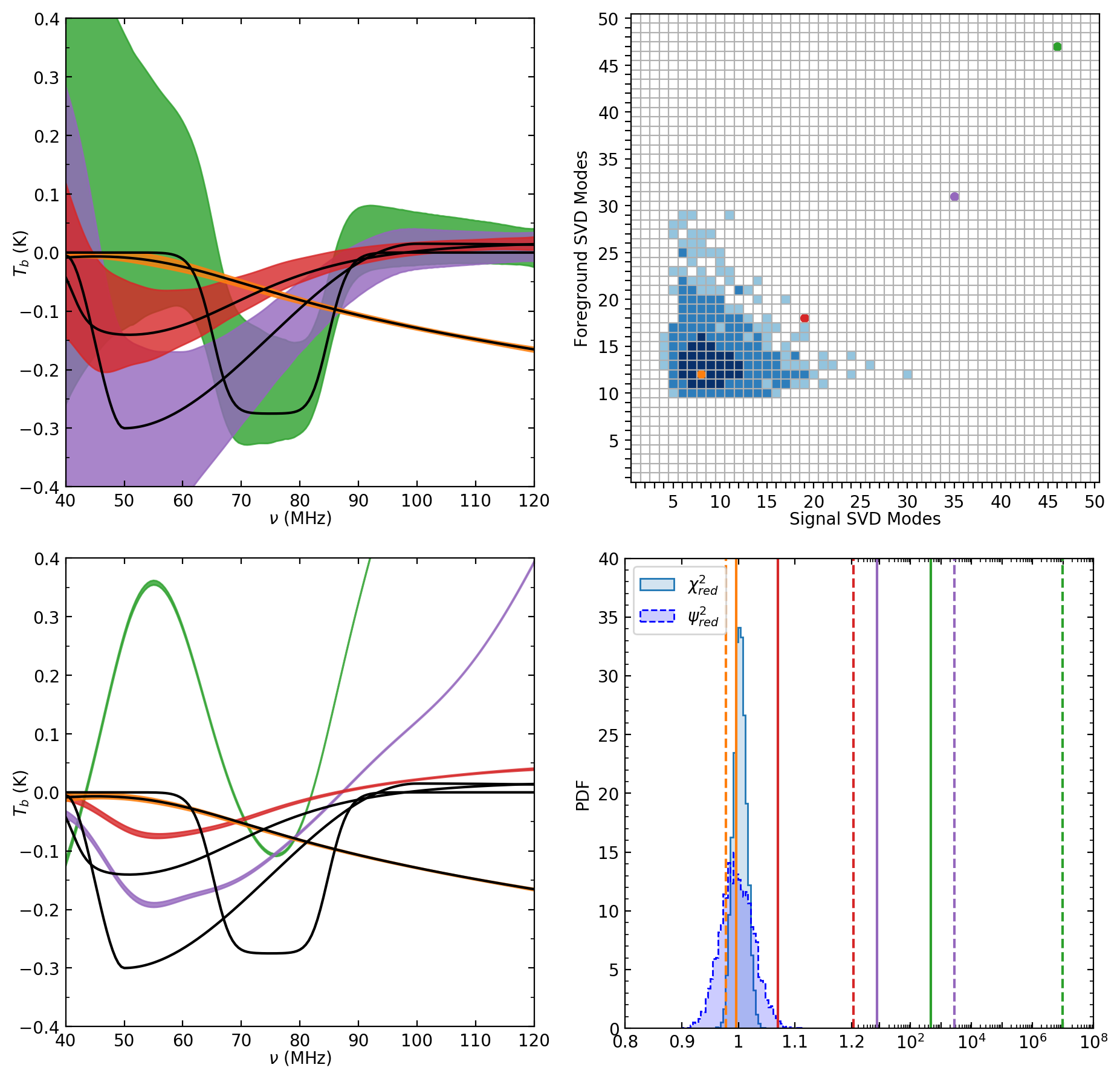}
    \caption{\textit{Left}: Input components used to create the simulated data. The orange, red, purple, and green beams and signals along with the black spectral index are used to create four realizations to be fit using the SVD modes from the training sets shown in blue. \textit{Middle}: 95\% confidence signal reconstructions produced by the pipeline for the four simulated data realizations both without (top) and with (bottom) training set priors. The color corresponds to the beam and signal in the left panel used to create the realization. \textit{Right}: Results of the tests without (top) and with (bottom) priors outlined in Section \ref{test_outline}. The top panel shows the SVD mode distribution when data realizations are made using training set curves (blue contours) as well as the numbers of modes chosen for the fits using the four example beams and signals (colored points). The blue contours enclose 68\%, 95\%, and 99.7\% of the 5000 fits performed. The bottom panel shows values of $\chi^2_{\textup{red}}$ (solid lines) and $\psi^2_{\textup{red}}$ (dashed lines) for fits including priors using the four example beams and signals compared to distributions of each statistic from 5000 data realizations made using training set curves. The x-axis scale is linear from $0.8 - 1.2$ and logarithmic above 1.2.}
    \label{fig:beam_and_signal}
\end{figure*}

Since the largest source of foreground emission is galactic synchrotron radiation \citep{Furlanetto:2006}, the foreground emission is well-modeled by a power law,
\begin{equation}
\label{power_law_equation}
    T_{sky}(\nu,\theta,\phi) = T_{map}(\theta,\phi)\Big(\frac{\nu}{\nu_0}\Big)^{\beta(\theta,\phi,\nu)},
\end{equation}
where $T_{map}$ is the foreground emission at the reference frequency $\nu_0$ and $\beta$ is the spectral index. Although the spectral index is often taken to be constant with a value of $\beta = -2.5$, in general $\beta$ may vary spatially and as a function of frequency (where the latter effect is typically referred to as spectral curvature). In fact, observations show that the spectral index flattens near the galactic plane (Hibbard et al. 2020; submitted to ApJ).

Note that although the foreground emission follows the form of equation \ref{power_law_equation}, the sky will be weighted by the antenna beam such that the measured antenna temperature is given by
\begin{equation}
    T_{ant}(\nu) = \frac{\int_{4\pi}T_{sky}(\nu,\theta,\phi)B(\nu,\theta,\phi)d\Omega}{\int_{4\pi}B(\nu,\theta,\phi)d\Omega},
\end{equation}
where $B$ is the antenna beam pattern, which in the case of the tests shown in this paper is assumed to be a Gaussian function of the zenith angle $\theta$ and will follow Equation \ref{gaussian_beam_equation}. Following Hibbard et al. 2020, we use two different analytical models for the spectral index as a function of galactic latitude: a constant and a Gaussian model. Due to the expected flattening of the spectral index near the galactic plane, we employ the Gaussian model to create the spectral index training set shown in the left middle panel of Figure \ref{fig:spectral_index}. The Gaussian spectral index model follows the equation
\begin{equation}
    \beta(\theta) = O + M\exp{\Big[\frac{-(\theta - \pi/2)^2}{2\sigma^2}\Big]},
\end{equation}
where $\theta$ is the galactic co-latitude and $\sigma$ roughly corresponds to the width of the galactic plane. For all of the training set curves, we assume the standard deviation to be $\sigma = 10^{\circ}$ while the offset ($O$) and magnitude of variation ($M$) parameters are drawn from the distributions $O \sim \mathcal{N}(-2.5, 0.005)$, $M \sim \mathcal{N}(-0.4, 0.01)$,\footnote{Note that the distribution for $M$ is truncated at 0.} where $\mathcal{N}(\mu, \sigma^2)$ denotes a normal distribution with mean $\mu$ and variance $\sigma^2$. In order to construct the full beam-weighted foreground training set on which to perform SVD, we interpolate the Haslam 408 MHz map to the relevant frequencies using the spectral index curves and convolve each of the resulting sky maps with the beam shown in the top left panel of Figure \ref{fig:spectral_index}.

For the examples to be fit, we first choose a spectral index realization directly from the training set as the fiducial example (shown in orange). We then choose the commonly used constant spectral index with $\beta = -2.5$ (shown in red) as well as two alternative Gaussian models with a smaller standard deviation of $\sigma = 5^{\circ}$ (shown in purple and green). Just as in Section \ref{beams}, we begin by performing the fits without using a prior distribution and construct the SVD mode histogram (upper middle and upper right panels of Figure \ref{fig:spectral_index}) before using the same data realizations but applying training set priors and calculating the $\chi^2_{\textup{red}}$ and $\psi^2_{\textup{red}}$ statistics (lower middle and lower right panels of Figure \ref{fig:spectral_index}).

\subsection{Signals}
\label{signals}

While the EDGES team claims a potential first detection of the 21-cm cosmic dawn signal \citep{Bowman:2018}, there is still significant controversy and therefore uncertainty regarding the true form of the 21-cm signal. For this reason, we choose to investigate several different parameterizations. We use the same signal training set as the examples outlined above in Sections \ref{beams} and \ref{spectral_index}. However, instead of creating data realizations with 21-cm signals taken from the training set, we input signals outside of the training set. To create these example signals, we employ several signal models, including a phenomenological parameterization known as the tanh model, a turning point model which interpolates between the local extrema of the 21-cm spectrum using a cubic spline, and the flattened Gaussian model of \cite{Bowman:2018}. For further details regarding the tanh and turning point models, see \cite{Harker:2016} and \cite{Rapetti:2019}, respectively. The same basic steps to perform the fits with and without priors are carried out as in Sections \ref{beams} and \ref{spectral_index} and the relevant plots are shown in Figure \ref{fig:signals}.

\subsection{Multiple Components}

While in the previous three subsections we have limited the mismatch between training set and data to a single component, it is possible that multiple components may have a deficiency. In order to investigate the effects of having multiple training sets that are insufficient, we create simulated data using the example beams from Section \ref{beams} and the example 21-cm signals from Section \ref{signals}. Beams and signals of the same color (i.e. orange, red, purple, or green) are used to create the data realizations. The signal extractions as well as the SVD mode distribution and the $\chi^2_{\textup{red}}$ and $\psi^2_{\textup{red}}$ statistics for these fits are shown in Figure \ref{fig:beam_and_signal}.

\section{Results and Discussion}
\label{results}

\subsection{Beams}
\label{beam_results}

\renewcommand{\arraystretch}{1.5}
\begin{table*}[t]
    \begin{tabular}{cccccccc}
    \multicolumn{8}{c}{No Priors}\\
    \hline
    \multirow{2}{*}{Example} & \multirow{2}{*}{$\chi_{\textup{red}}^2$} & \multirow{2}{*}{$\psi^2_{\textup{red}}$} & \# signal & \# foreground & \# modes & RMS signal & RMS signal\\
     & & & modes & modes & p-value & uncertainty [mK] & bias [mK]\\
    \hline\hline
    \multicolumn{8}{c}{Beams (Figure \ref{fig:beams})}\\
    orange & 1.00 & 0.99 & 9 & 13 & 0.640 & 3.6 & 0.8\\
    red & 1.00 & 0.99 & 8 & 21 & 0.046 & 3.3 & 1.0\\
    purple & 1.00 & 0.99 & 11 & 30 & 0 & 31 & 36\\
    green & 1.02 & 1.04 & 19 & 48 & 0 & 60. & 74\\
    \hline
    \multicolumn{8}{c}{Spectral Index Models (Figure \ref{fig:spectral_index})}\\
    orange & 1.00 & 0.98 & 8 & 10 & 0.874 & 2.1 & 1.4\\
    red & 1.00 & 0.99 & 9 & 14 & 0.126 & 6.2 & 1.0\\
    purple & 1.01 & 1.01 & 6 & 32 & 0 & 23 & 4.4\\
    green & 1.04 & 1.09 & 7 & 47 & 0 & 48 & 35\\
    \hline
    \multicolumn{8}{c}{21-cm Signals (Figure \ref{fig:signals})}\\
    orange & 0.99 & 0.98 & 8 & 12 & 0.950 & 2.3 & 0.9\\
    red & 0.99 & 1.00 & 19 & 13 & 0.033 & 10. & 2.6\\
    purple & 1.02 & 1.04 & 41 & 12 & 0 & 10. & 4.1\\
    green & 0.99 & 0.97 & 46 & 12 & 0 & 10. & 4.8\\
    \hline
    \multicolumn{8}{c}{Beams and 21-cm Signals (Figure \ref{fig:beam_and_signal})}\\
    orange & 0.99 & 0.98 & 8 & 12 & 0.950 & 2.3 & 0.9\\
    red & 0.99 & 1.00 & 19 & 18 & 0 & 19 & 13\\
    purple & 1.02 & 1.04 & 35 & 31 & 0 & 54 & 30.\\
    green & 1.00 & 0.98 & 46 & 47 & 0 & 67 & 74\\
    \hline
    \end{tabular}
    \caption{Statistics regarding the goodness-of-fit and the 21-cm signal extraction for the examples shown in Figures \ref{fig:beams}, \ref{fig:spectral_index}, and \ref{fig:signals} without a prior distribution. For a description of how the ``\# modes p-value'' column is calculated, see footnote \ref{svd_modes_percentile_footnote}. RMS signal uncertainty refers to the root-mean-square of the 68\% confidence interval on the 21-cm signal, while RMS signal bias indicates the root-mean-square of the residual between the maximum likelihood reconstruction of the 21-cm signal and the input 21-cm signal.}
    \label{tab:no_priors_table}
\end{table*}

We begin by discussing the fits performed without priors shown in the upper panels of Figure \ref{fig:beams}. Additional information about the quality of the 21-cm signal extraction for these fits performed without priors is provided in Table \ref{tab:no_priors_table}. Columns 2 and 3 of the table show the $\chi^2_{\textup{red}}$ and $\psi^2_{\textup{red}}$ values for the fits, indicating that the full data is fit well in all of the examples. The ``\# modes p-value'' is a measure of the portion of the training set SVD mode distribution that lies outside a given fit.\footnote{The p-value is calculated by sorting the SVD mode number grid by the number of training set realizations within each grid square and performing a cumulative sum until the grid square chosen for the given fit is reached.  The remaining number of realizations outside of the given fit is then divided by the 5000 total realizations performed to find the p-value.\label{svd_modes_percentile_footnote}} A p-value of 0 indicates that the fit lies entirely outside of the training set distribution. The value is exactly 0 rather than some small non-zero number due to the discrete nature of the distribution. For the orange example, the p-value is large, meaning that it is consistent with the number of SVD modes chosen for the training set data realizations. This is not surprising since the orange beam is well contained within the training set distribution. However, as the beams become more different from the training set and the number of required foreground SVD modes becomes larger, the fits move further from the training set distribution. Using 0.05 as our significance threshold, we would conclude that in the red, purple, and green examples (all of which use beams outside of the training set) the training sets are insufficient based on their p-values.

As the fits move further from the reference distribution in the SVD mode space, the size of the uncertainty band on the reconstructed signal and the bias between the input and the reconstruction of the signal (as shown visually in the upper middle panel of Figure \ref{beams} and quantitatively by the final two columns of Table \ref{tab:no_priors_table}) grow. Note, however, that the increases in uncertainty and bias are approximately proportional such that the ratio between the bias and the uncertainty is roughly constant. In these examples it seems that the number of SVD modes can be used as a proxy for the similarity between the input component and the training set. This relationship will be explored further in the following subsections.

Now that we have performed the linear fit without priors, we may wish to apply priors derived from the training sets, as described in Section \ref{priors}. The increased constraints due to the priors require all components of the data to be modeled sufficiently by the training sets in order for the full data to be fit well. Thus, the power of $\chi^2_{\textup{red}}$ and $\psi^2_{\textup{red}}$ is increased when priors are used.

The lower panels of Figure \ref{fig:beams} show the signal reconstructions and the $\chi^2_{\textup{red}}$ and $\psi^2_{\textup{red}}$ values for the fits performed with priors. Additional information for the fits is provided in Table \ref{tab:priors_table}. Similar to the test without priors, the orange example, which is drawn from the same distribution as the training set, lies well within the distributions for both statistics. For the remaining examples, the further the beam lies from the training set, the more biased the signal reconstructions become and the more divergent the values of the statistics are from the training set distributions. Again, the red, purple, and green examples are excluded by the 0.05 significance threshold for the p-value using both $\chi^2$ and $\psi^2$. That the signal reconstructions for these examples are inaccurate is reasonable due to the fact that the prior distribution is constraining the foreground to models than cannot describe these realizations.

The red example, in which the beam is near the edge of the training set, is particularly interesting. The $\chi^2_{\textup{red}}$ value, while close to the edge of the training set $\chi^2_{\textup{red}}$ distribution, is still contained within it. This can be readily seen by the fact that the $\chi^2_{\textup{red}}$ p-value is greater than 0. The $\psi^2_{\textup{red}}$ value, however, is clearly separated from the training set $\psi^2_{\textup{red}}$ distribution with a $\psi^2_{\textup{red}}$ p-value of 0. When we look at the quality of the signal reconstruction for the fits with priors in the lower middle panel of Figure \ref{fig:beams} and the final two columns of Table \ref{tab:priors_table}, the signal bias is significantly larger than the 68\% uncertainty band, indicating that the uncertainties are spuriously small. This example begins to demonstrate the utility of $\psi^2$ in recognizing poor signal extractions that $\chi^2$ may be unable to identify.

For the purple and green examples, both $\chi^2_{\textup{red}}$ and $\psi^2_{\textup{red}}$ clearly indicate a poor fit to the data and the signal bias increases significantly for both examples as expected. Note that the spuriously small uncertainty intervals on the signal extractions for these fits are a product of the limited covariance of the posterior enforced by the prior distribution. Excluding the orange beam, these examples are true ``failures'' (caused by an insufficient training set) in the sense that the signal reconstruction is truly inaccurate and not simply imprecise. 

\subsection{Spectral Indices}
\label{spectral_index_results}

\begin{table*}
    \centering
    \begin{tabular}{ccccccc}
    \multicolumn{7}{c}{Priors}\\
    \hline
    \multirow{2}{*}{Example} & \multirow{2}{*}{$\chi_{\textup{red}}^2$} & $\chi_{\textup{red}}^2$ & \multirow{2}{*}{$\psi^2_{\textup{red}}$} & $\psi^2_{\textup{red}}$ & RMS signal & RMS signal\\
     & & p-value & & p-value & uncertainty [mK] & bias [mK]\\
    \hline\hline
    \multicolumn{7}{c}{Beams (Figure \ref{fig:beams})}\\
    orange & 1.00 & 0.849 & 0.99 & 0.820 & 1.6 & 1.3 \\
    red & 1.04 & 0.001 & 1.20 & 0 & 1.6 & 7.7 \\
    purple & 7.87 & 0 & $2.68 \times 10^3$ & 0 & 1.6 & 121 \\
    green & 476 & 0 & $9.80 \times 10^6$ & 0 & 1.6 & 853 \\
    \hline
    \multicolumn{7}{c}{Spectral Index Models (Figure \ref{fig:spectral_index})}\\
    orange & 0.99 & 0.268 & 0.96 & 0.189 & 1.6 & 0.9 \\
    red & 1.01 & 0.523 & 1.00 & 0.952 & 2.0 & 2.6 \\
    purple & 152 & 0 & $2.45 \times 10^6$ & 0 & 1.9 & 420 \\
    green & 311 & 0 & $1.06 \times 10^7$ & 0 & 2.3 & 850 \\
    \hline
    \multicolumn{7}{c}{21-cm Signals (Figure \ref{fig:signals})}\\
    orange & 1.00 & 0.849 & 0.98 & 0.549 & 1.6 & 0.6 \\
    red & 1.03 & 0.011 & 1.12 & 0 & 1.6 & 44 \\
    purple & 1.19 & 0 & 4.20 & 0 & 1.6 & 42 \\
    green & 2.72 & 0 & 761 & 0 & 1.6 & 13 \\
    \hline
    \multicolumn{7}{c}{Beams and 21-cm Signals (Figure \ref{fig:beam_and_signal})}\\
    orange & 1.00 & 0.849 & 0.98 & 0.549 & 1.6 & 0.6\\
    red & 1.07 & 0 & 1.36 & 0 & 1.6 & 45\\
    purple & 8.22 & 0 & $2.79 \times 10^3$ & 0 & 1.6 & 140\\
    green & 478 & 0 & $9.85 \times 10^6$ & 0 & 1.6 & 850\\
    \hline
    \end{tabular}
    \caption{Statistics regarding the goodness-of-fit for the full data and the 21-cm signal extraction for the examples shown in Figures \ref{fig:beams}, \ref{fig:spectral_index}, and \ref{fig:signals}. Fits are performed using a prior distribution for all components derived from the training set curves. The $\chi_{\textup{red}}^2$ and $\psi_{\textup{red}}^2$ p-value columns are the percentage of training set realizations with values of $\chi_{\textup{red}}^2$ and $\psi_{\textup{red}}^2$ more extreme than the corresponding values for the given fit. RMS signal uncertainty refers to the root-mean-square of the 68\% confidence interval on the 21-cm signal, while RMS signal bias indicates the root-mean-square of the residual between the maximum likelihood reconstruction of the 21-cm signal and the input 21-cm signal.}
    \label{tab:priors_table}
\end{table*}

An inspection of the upper middle panel of Figure \ref{fig:spectral_index} shows that for the fits without priors, the signal is extracted quite well from the orange example taken from the spectral index training set, as is expected. Perhaps more unexpected, however, is that the constant spectral index data realization, plotted in red, is also extracted quite well, despite the spectral index following a different model than the training set curves. Although the fit does not lie right in the center of the reference SVD mode distribution, it still lies within this distribution and the p-value is indeed greater than the 0.05 significance threshold. While the training set only includes Gaussian spectral index realizations, some of the curves within the training set have small magnitudes of variation, making them flat enough that a constant model realization can be fit well with the modes from the Gaussian training set. The increased sharpness near the galactic plane of the purple and green examples, however, requires many more foreground SVD modes to fit them well. Similar to the examples shown in Section \ref{beam_results}, the signal extractions for the realizations that require more foreground modes than the training set distribution are much less precise while maintaining a similar level of accuracy.

Even when including priors, both the orange and red realizations can still be fit well, as shown by the $\chi^2_{\textup{red}}$ and $\psi^2_{\textup{red}}$ values in the lower right panel of Figure \ref{fig:spectral_index}, and the signal extraction is more precise than the fit without priors, as shown in the lower middle panel. However, the two thinner Gaussian examples in purple and green have large values for both $\chi^2_{\textup{red}}$ and $\psi^2_{\textup{red}}$ compared to the training set distributions (corresponding to p-values of 0) and wildly inaccurate signal reconstructions.

\subsection{Signals}
\label{signal_results}

In Sections \ref{beams} and \ref{spectral_index} the foreground input was inconsistent with the foreground training set, but in Section \ref{signals} we use input 21-cm signals that are inconsistent with the signal training set. With these examples we find that the number of foreground modes is consistent with the bulk of the training set realizations, but some of the input signals, the turning point and flattened Gaussian models in particular, require additional signal modes. The fact that changes to the foreground component affect the number of foreground SVD modes and changes to the signal component affect the number of signal SVD modes can provide extremely useful information when analyzing real observations. If many more signal SVD modes are required while the number of foreground modes is consistent with the training set distribution, it is likely that the signal training set, rather than the foreground training set, needs to be altered. This can save time and effort when a poor fit is detected. However, it is important to note that this distinction may not be as clear for all foreground and signal model combinations. If additional overlap between the two models exists (for example, in an analysis of a single spectrum without polarization measurements), this bimodality may be less evident, which could lead to confusion regarding which training set(s) may need to be altered. Further investigation of the effect of increased overlap is left to future work (see, however, Section~\ref{multiple_components_results} for additional insights).

Even though a larger number of signal modes are required for some examples, the signal extractions for the fits without priors are still relatively precise compared to the least precise examples of Section \ref{beams} and \ref{spectral_index}. However, when using training set priors, the lower middle panel of Figure \ref{fig:signals} shows that the red, purple, and green examples (all of which are from models not included in the signal training set) cannot be fit accurately. Even if the signal extraction produced by the linear fit without priors is deemed to be sufficient, care should be taken before moving forward with the analysis. For example, after performing the linear fit, one may wish to run a Markov Chain Monte Carlo (MCMC) to estimate nonlinear parameter values \citep[see][]{Rapetti:2019}, which requires an adequate model for the signal.

\subsection{Multiple Components}
\label{multiple_components_results}
Instead of limiting the difference between the training set and the simulated data to a single component, in Figure \ref{fig:beam_and_signal} both the beam and the signal are varied relative to the training sets. An inspection of the SVD mode grid in these cases shows that the fits tend to lie along a diagonal line, requiring an additional number of both foreground and signal modes compared to the training set distribution. Contrast this with Figures \ref{fig:beams}, \ref{fig:spectral_index}, and \ref{fig:signals} in which the fits generally require more modes for only the component that lies outside of its training set. In the case that priors are used, the $\chi^2_{\textup{red}}$ and $\psi^2_{\textup{red}}$ values indicate poor fits for the red, purple, and green cases, but any information that might indicate that multiple training sets are insufficient is lost.

Although at first glance it would seem that a fit that requires more modes for multiple components compared to the training set distribution would indicate that both training sets are insufficient in some way (as is the case for the examples in Figure \ref{fig:beam_and_signal}), it is possible that substantial overlap between the foreground and signal models could lead to a similar result when only one training set is deficient. The simulated observations used in this paper utilized polarization as well as multiple pointings and rotation angles to decrease overlap (and, hence, covariances) between the components. However, other observation strategies such as single spectrum experiments (analogous to the EDGES) will lead to greater overlap between the foreground and signal components.

Thus, there are two possibilities if additional SVD modes are required for multiple components:
\begin{enumerate}
    \item Multiple training sets are insufficient \textit{or}
    \item Only one training set is insufficient, but a significant degeneracy between the components exists.
\end{enumerate}
We propose two avenues with which to determine which of the above is true for a given fit. First, one may compute the overlap matrix $\boldsymbol{D}$ between the components. If we suppose that $\boldsymbol{F}_{\textup{fg}}$ and $\boldsymbol{F}_{21}$ are matrices containing the SVD modes for the foreground and 21-cm signal, respectively, and that $\boldsymbol{\Psi}_{21}$ is the signal expansion matrix which encodes how the signal is found in the full data (i.e. only in Stokes I), the overlap matrix is given by
\begin{equation}
    \boldsymbol{D} = \boldsymbol{F}_{\textup{fg}}^T\boldsymbol{C}^{-1}\boldsymbol{\Psi}_{21}\boldsymbol{F}_{21}.
\end{equation}
In order to form a summary statistic for the overlap, we can calculate the normalized RMS error on the signal (NRMS$_{21}$), which is the RMS of the ratio between the $1\sigma$ uncertainty on the signal and the $1\sigma$ noise level on the data. This normalized error can be expressed as a sum over the eigenvalues $\lambda_j$ of $\boldsymbol{D}^T\boldsymbol{D}$:
\begin{equation}
    \textup{NRMS}_{21} = \sqrt{\frac{1}{n_{\nu}}\sum_{j=1}^{n_{21}}\frac{1}{1 - \lambda_j}},
\end{equation}
where $n_{\nu}$ is the number of frequencies and $n_{21}$ is the number of modes chosen for the 21-cm signal component.\footnote{For a full derivation see section 2.2 of \cite{Tauscher:2020a}.} If the foreground and signal bases are entirely orthogonal, then the eigenvalues of $\boldsymbol{D}$ are all 0 and NRMS$_{21}$ is minimized. if there is an exact degeneracy (i.e. one of the foreground modes can be expressed as a linear combination of signal modes or vice versa), then one or more of the eigenvalues of $\boldsymbol{D}$ will be exactly 1 and NRMS$_{21}$ goes to $\infty$. In general, the eigenvalues will lie between these extremes. Although it is unclear what eigenvalues or what value of NRMS$_{21}$ consititutes ``significant overlap'' between the components, examination of these quantities may provide insight. Further study of how the overlap between components may affect the chosen number of modes is left to future work.

An alternative method of investigation is through the Minimum Assumption Analysis (MAA) introduced in \cite{Tauscher:2020b}. Rather than imposing a model on the signal through a training set, the MAA makes no assumptions about the spectral shape of the signal, only that the signal appears the same in all total power spectra. By applying the MAA, we can effectively marginalize the foreground model out of the data. If the foreground training set is sufficient, then between the foreground modes and the MAA signal, the full data should be able to be fit well down to the noise level. If this is not the case (for example, if there is some component of the data that cannot be fit by the foreground modes but does not appear the same in every spectrum), then it is likely that the foreground training set is insufficient. As with the overlap matrix, we leave a demonstration of this method in practice to future work.

\section{Conclusions}
\label{conclusions}

When applied to real observations, how do we know if training sets of 21-cm signals and beam-averaged foregrounds are sufficient to analyze observations both accurately and precisely? The goal of this work is to devise a method to detect cases in which one or more components of a given data realization may be inadequately described by their respective training set, thus ensuring the robustness of the analysis performed with our pipeline. This is especially crucial when analyzing real low frequency observations where the true form of each component, particularly the 21-cm signal, is unknown. When no prior distribution is used for the fit, commonly used goodness-of-fit statistics such as $\chi^2_{\textup{red}}$, or the recently introduced $\psi^2_{\textup{red}}$, are insufficient to serve this purpose due to the fact that they are only able to assess the fit to the full data. When modeling multiple components simultaneously, it is possible for overlap (i.e. covariances) between the SVD models of the components to lead to suboptimal extraction of the 21-cm signal even when the full data is fit well.

When no prior distribution is used for any of the components, we find that the number of SVD modes chosen for each component of the data is able to provide information about the relationship between the input component and the training set. The numbers of modes chosen for a given data realization must be compared to a distribution created by fitting many realizations of data created from curves taken directly from the training sets. If the numbers of SVD modes chosen for a given fit is inconsistent with this training set distribution, we find that it is likely that the training sets do not adequately describe the data. Furthermore, we find that the location of the fit in this SVD mode space relative to the training set distribution can be used to determine which training set needs to be altered. Although we note that foreground and signal models with additional overlap may lead to increased mixing between the modes of the components, all of the cases presented in this work, which are based on realistic signal and beam-weighted foreground modeling, show that if the foreground training set is inadequate, the number of foreground SVD modes increases while leaving the number of signal SVD modes relatively unaffected and vice versa. If a poor fit is detected, rather than attempting to improve both training sets through trial and error, this test is able to pinpoint the component that requires attention. While the simulated observations used here contain only two non-noise components, this same method can be straightforwardly extended to data with a larger number of components.

We also explore a similar test for fits employing a prior distribution on the SVD mode coefficients derived from the training sets. When including a training set prior, if the training sets do not fully encompass the true form of each component the full data will not be able to be fit well, making the $\chi^2_{\textup{red}}$ and $\psi^2_{\textup{red}}$ statistics better able to discern when one or more of the training sets may be inadequate. After calculating the values of these statistics for a given fit we must again compare them to a reference distribution of the statistic values formed by fitting many data realizations made from the training set curves. If the statistic value for a given fit is not consistent with the reference distribution to within the desired threshold, the training sets are likely inconsistent with the observed data. We find that in some cases, for example when the true form of a component lies near the edge of the distribution of training set curves, $\psi^2_{\textup{red}}$ is able to better detect this discrepancy.

It may be possible for a given combination of data and training sets to ``pass'' the test without priors, but when a training set prior is applied the fit is no longer consistent with the reference distribution. One possible way in which this could occur is if the shapes of the SVD modes derived from the training set are sufficient to describe the data but there is a discrepancy between the magnitudes of the components in the training sets and the data. Without a prior distribution, since the SVD mode coefficients are allowed to take any value, the magnitudes of the training set curves are irrelevant. When training set priors are added, however, the full distribution of the curves including their shape and magnitude is taken into account. Including priors therefore leads to a much more stringent test of the ability of the training sets to describe the data, so care should be taken to ensure that the training sets are truly representative of both the shapes and magnitudes of the components before including priors in the fit.

The steps outlined in this paper describe tests that can be performed when analyzing any data with our pipeline. The tests provide insight as to whether the modeling employed is representative of the data, lending reliability to results from our training set based framework for 21-cm global signal analysis and should thus be extremely valuable for current and upcoming low frequency experiments.

\section{Acknowledgements}
This work is directly supported by the NASA Solar System Exploration Research Virtual Institute cooperative agreement number 80ARC017M0006. DR was supported for part of the time by a NASA Postdoctoral Program Senior Fellowship at NASA's Ames Research Center, administered by the Universities Space Research Association under contract with the National Aeronautics and Space Administration (NASA). This work was also supported by NASA under award number NNA16BD14C for NASA Academic Mission Services.

\bibliographystyle{yahapj}
\bibliography{ref}

\providecommand{\noopsort}[1]{}
\begin{thebibliography}{}
\providecommand\natexlab[1]{#1}
\providecommand\JournalTitle[1]{#1}

\bibitem[{{Barkana}(2018)}]{Barkana:2018}
{Barkana}, R. 2018,
  \href{http://dx.doi.org/10.1038/nature25791}{\JournalTitle{Nature}, 555, 71}

\bibitem[{Bassett {et~al.}(2020)Bassett, Rapetti, Burns, Tauscher, \&
  MacDowall}]{Bassett:2020}
Bassett, N., Rapetti, D., Burns, J.~O., Tauscher, K., \& MacDowall, R. 2020,
  \href{http://dx.doi.org/https://doi.org/10.1016/j.asr.2020.05.050}{\JournalTitle{Advances
  in Space Research}, 66, 1265 }

\bibitem[{{Berlin} {et~al.}(2018){Berlin}, {Hooper}, {Krnjaic}, \&
  {McDermott}}]{Berlin:2018}
{Berlin}, A., {Hooper}, D., {Krnjaic}, G., \& {McDermott}, S.~D. 2018,
  \href{http://dx.doi.org/10.1103/PhysRevLett.121.011102}{\JournalTitle{Physical
  Review Letters}, 121, 011102}

\bibitem[{{Bowman} {et~al.}(2018){Bowman}, {Rogers}, {Monsalve}, {Mozdzen}, \&
  {Mahesh}}]{Bowman:2018}
{Bowman}, J.~D., {Rogers}, A. E.~E., {Monsalve}, R.~A., {Mozdzen}, T.~J., \&
  {Mahesh}, N. 2018,
  \href{http://dx.doi.org/10.1038/nature25792}{\JournalTitle{Nature}, 555, 67}

\bibitem[{{Bradley} {et~al.}(2019){Bradley}, {Tauscher}, {Rapetti}, \&
  {Burns}}]{Bradley:2019}
{Bradley}, R.~F., {Tauscher}, K., {Rapetti}, D., \& {Burns}, J.~O. 2019,
  \href{http://dx.doi.org/10.3847/1538-4357/ab0d8b}{\JournalTitle{\apj}, 874,
  153}

\bibitem[{{Burns}(2020)}]{Burns:2020}
{Burns}, J.~O. 2020, \JournalTitle{arXiv e-prints}, arXiv:2003.06881

\bibitem[{{Ewall-Wice} {et~al.}(2018){Ewall-Wice}, {Chang}, {Lazio},
  {Dor{\'e}}, {Seiffert}, \& {Monsalve}}]{Ewall-Wice:2018}
{Ewall-Wice}, A., {Chang}, T.~C., {Lazio}, J., {et~al.} 2018,
  \href{http://dx.doi.org/10.3847/1538-4357/aae51d}{\JournalTitle{\apj}, 868,
  63}

\bibitem[{{Feng} \& {Holder}(2018)}]{Feng&Holder:2019}
{Feng}, C. \& {Holder}, G. 2018,
  \href{http://dx.doi.org/10.3847/2041-8213/aac0fe}{\JournalTitle{Astrophysical
  Journal Letters}, 858, L17}

\bibitem[{{Fialkov} \& {Barkana}(2019)}]{Fialkov&Barkana:2019}
{Fialkov}, A. \& {Barkana}, R. 2019,
  \href{http://dx.doi.org/10.1093/mnras/stz873}{\JournalTitle{Monthly Notices
  of the Royal Astronomical Society}, 486, 1763}

\bibitem[{{Fialkov} {et~al.}(2018){Fialkov}, {Barkana}, \&
  {Cohen}}]{Fialkov:2018}
{Fialkov}, A., {Barkana}, R., \& {Cohen}, A. 2018,
  \href{http://dx.doi.org/10.1103/PhysRevLett.121.011101}{\JournalTitle{Physical
  Review Letters}, 121, 011101}

\bibitem[{{Furlanetto} {et~al.}(2006){Furlanetto}, {Oh}, \&
  {Briggs}}]{Furlanetto:2006}
{Furlanetto}, S.~R., {Oh}, S.~P., \& {Briggs}, F.~H. 2006,
  \href{http://dx.doi.org/10.1016/j.physrep.2006.08.002}{\JournalTitle{Physics
  Reports}, 433, 181}

\bibitem[{{Harker} {et~al.}(2016){Harker}, {Mirocha}, {Burns}, \&
  {Pritchard}}]{Harker:2016}
{Harker}, G. J.~A., {Mirocha}, J., {Burns}, J.~O., \& {Pritchard}, J.~R. 2016,
  \href{http://dx.doi.org/10.1093/mnras/stv2630}{\JournalTitle{\mnras}, 455,
  3829}

\bibitem[{{Haslam} {et~al.}(1982){Haslam}, {Salter}, {Stoffel}, \&
  {Wilson}}]{Haslam:1982}
{Haslam}, C.~G.~T., {Salter}, C.~J., {Stoffel}, H., \& {Wilson}, W.~E. 1982,
  \JournalTitle{Astronomy an Astrophysics}, 47, 1

\bibitem[{{Hills} {et~al.}(2018){Hills}, {Kulkarni}, {Meerburg}, \&
  {Puchwein}}]{Hills:2018}
{Hills}, R., {Kulkarni}, G., {Meerburg}, P.~D., \& {Puchwein}, E. 2018,
  \href{http://dx.doi.org/10.1038/s41586-018-0796-5}{\JournalTitle{\nat}, 564,
  E32}

\bibitem[{{Loeb} \& {Mu{\~n}oz}(2018)}]{Loeb&Munoz:2018}
{Loeb}, A. \& {Mu{\~n}oz}, J.~B. 2018,
  \href{http://dx.doi.org/10.1103/Physics.11.69}{\JournalTitle{Physics Online
  Journal}, 11, 69}

\bibitem[{{Mebane} {et~al.}(2020){Mebane}, {Mirocha}, \&
  {Furlanetto}}]{Mebane:2020}
{Mebane}, R.~H., {Mirocha}, J., \& {Furlanetto}, S.~R. 2020,
  \href{http://dx.doi.org/10.1093/mnras/staa280}{\JournalTitle{\mnras}, 493,
  1217}

\bibitem[{{Mirocha}(2014)}]{Mirocha:2014}
{Mirocha}, J. 2014,
  \href{http://dx.doi.org/10.1093/mnras/stu1193}{\JournalTitle{\mnras}, 443,
  1211}

\bibitem[{{Nhan} {et~al.}(2019){Nhan}, {Bordenave}, {Bradley}, {Burns},
  {Tauscher}, {Rapetti}, \& {Klima}}]{Nhan:2019}
{Nhan}, B.~D., {Bordenave}, D.~D., {Bradley}, R.~F., {et~al.} 2019,
  \href{http://dx.doi.org/10.3847/1538-4357/ab391b}{\JournalTitle{\apj}, 883,
  126}

\bibitem[{{Pritchard} \& {Loeb}(2012)}]{Pritchard:2012}
{Pritchard}, J.~R. \& {Loeb}, A. 2012,
  \href{http://dx.doi.org/10.1088/0034-4885/75/8/086901}{\JournalTitle{Reports
  on Progress in Physics}, 75, 086901}

\bibitem[{{Rapetti} {et~al.}(2020){Rapetti}, {Tauscher}, {Mirocha}, \&
  {Burns}}]{Rapetti:2019}
{Rapetti}, D., {Tauscher}, K., {Mirocha}, J., \& {Burns}, J.~O. 2020,
  \href{http://dx.doi.org/10.3847/1538-4357/ab9b29}{\JournalTitle{\apj}, 897,
  174}

\bibitem[{{Sims} \& {Pober}(2020)}]{Sims&Pober:2020}
{Sims}, P.~H. \& {Pober}, J.~C. 2020,
  \href{http://dx.doi.org/10.1093/mnras/stz3388}{\JournalTitle{\mnras}, 492,
  22}

\bibitem[{{Spinelli} {et~al.}(2019){Spinelli}, {Bernardi}, \&
  {Santos}}]{Spinelli:2019}
{Spinelli}, M., {Bernardi}, G., \& {Santos}, M.~G. 2019,
  \href{http://dx.doi.org/10.1093/mnras/stz2425}{\JournalTitle{\mnras}, 489,
  4007}

\bibitem[{{Tauscher} {et~al.}(2018{\natexlab{a}}){Tauscher},
  {\noopsort{aa}}{Rapetti}, {Burns}, \& {Switzer}}]{Tauscher:2018}
{Tauscher}, K., {\noopsort{aa}}{Rapetti}, D., {Burns}, J.~O., \& {Switzer}, E.
  2018{\natexlab{a}},
  \href{http://dx.doi.org/10.3847/1538-4357/aaa41f}{\JournalTitle{Astrophysical
  Journal}, 853, 187}

\bibitem[{{Tauscher} {et~al.}(2018{\natexlab{b}}){Tauscher},
  {\noopsort{ab}}{Rapetti}, \& {Burns}}]{Tauscher:2018b}
{Tauscher}, K., {\noopsort{ab}}{Rapetti}, D., \& {Burns}, J.~O.
  2018{\natexlab{b}},
  \href{http://dx.doi.org/10.1088/1475-7516/2018/12/015}{\JournalTitle{Journal
  of Cosmology and Astroparticle Physics}, 2018, 015}

\bibitem[{{Tauscher} {et~al.}(2020{\natexlab{a}}){Tauscher},
  {\noopsort{ba}}{Rapetti}, \& {Burns}}]{Tauscher:2020a}
{Tauscher}, K., {\noopsort{ba}}{Rapetti}, D., \& {Burns}, J.~O.
  2020{\natexlab{a}},
  \href{http://dx.doi.org/10.3847/1538-4357/ab9b2a}{\JournalTitle{\apj}, 897,
  175}

\bibitem[{{Tauscher} {et~al.}(2020{\natexlab{b}}){Tauscher},
  {\noopsort{bb}}{Rapetti}, \& {Burns}}]{Tauscher:2020b}
{Tauscher}, K., {\noopsort{bb}}{Rapetti}, D., \& {Burns}, J.~O.
  2020{\natexlab{b}},
  \href{http://dx.doi.org/10.3847/1538-4357/ab9a3f}{\JournalTitle{\apj}, 897,
  132}

\end{thebibliography}

\end{document}